\documentclass{PoS}

\usepackage{amsmath}
\usepackage{amssymb}
\usepackage{graphicx}
\usepackage{epstopdf}
\usepackage{siunitx}
\usepackage{multicol}
\usepackage{verbatim}

\newcommand{\ms}{\overline{MS}}

\title{Recent Developments in $x$-dependent Structure Calculations}

\ShortTitle{$x$-dependent Structure}

\author{\speaker{Christopher J.~Monahan}\\
Institute for Nuclear Theory, University of Washington,
Seattle, Washington 98195-1550, USA\\
        E-mail: \email{cjm373@uw.edu}}

\abstract{First principles calculations of the Bjorken-$x$ dependence of hadron structure have been a long-standing challenge for lattice QCD. This year marks a significant milestone: the first determinations of parton distribution functions, which capture the longitudinal momentum structure of fast-moving hadrons, at physical pion masses. Moreover, there has been significant progress in our understanding of the theoretical frameworks underpinning these calculations, although not all sources of systematic uncertainty have been fully explored. I review the various approaches to extracting $x$-dependent hadron structure from lattice QCD and highlight recent results in both the meson and baryon sectors.}

\FullConference{The 36th Annual International Symposium on Lattice Field Theory - LATTICE2018\\
		22-28 July, 2018\\
		Michigan State University, East Lansing, Michigan, USA.}

\begin{document}
\setlength{\abovedisplayskip}{0.5\baselineskip}
\setlength{\belowdisplayskip}{0.5\baselineskip}

\section{\label{sec:intro}Introduction}

Protons and neutrons, collectively referred to as nucleons, are the basic building blocks of the visible Universe. Far from being simple building blocks, however, nucleons are complicated, strongly-coupled dynamical systems with rich internal structure that is poorly understood. Our ignorance of nucleon structure translates directly into uncertainty in several fundamental parameters of the Standard Model of particle physics. For example, the uncertainties in parton distribution functions (PDFs), which describe the structure of nucleons in terms of the longitudinal momentum fraction carried by their constituent quarks and gluons (or partons), are the dominant, or one of the dominant, theoretical uncertainties in Higgs couplings and the mass of the $W$ boson \cite{Gao:2017yyd}. 

Until recently, attempts to directly calculate PDFs have been caught on the horns of a dilemma: on the one horn, PDFs are defined in terms of nonperturbative matrix elements of fields at light-like separations, for which no systematic nonperturbative techniques exist. On the other, lattice quantum chromodynamics (QCD)--the only \emph{ab initio} nonperturbative approach to QCD--is formulated in Euclidean spacetime, precluding the direct calculation of light cone physics. Here I review the theoretical developments that have overcome, or circumvented, these twin challenges, leading to the first lattice calculations of PDFs at physical pion masses, and highlight some recent results. As part of this review, I identify three emerging, interwoven themes: systematic uncertainties; theoretical challenges; and the future role of precision lattice calculations.

I start with a brief overview of phenomenological extractions of PDFs. In Sec.~\ref{sec:matelem} I review the quasi PDF and pseudo PDF approaches, focussing on the most recent developments. I then turn to other first-principles approaches to $x$-dependent structure, starting with factorisable matrix elements in Sec.~\ref{sec:facmat}. In Sec.~\ref{sec:hadront} I discuss calculations of the Euclidean hadronic tensor and Compton amplitude, before touching on two approaches to calculating multiple Mellin moments in Sec.~\ref{sec:many}.

\section{\label{sec:pdfs} PDFs from global fits}

PDFs characterise the structure of hadrons in terms of their constituent quarks and gluons and arise in the hard-scattering factorisation of inclusive QCD processes, such as deep inelastic scattering (DIS). A complete review of factorisation is beyond the scope of this review (see, for example, \cite{Collins:2011zzd}), but I lay the groundwork for my discussion by introducing the determination of PDFs from global fits to experimental data. For more thorough reviews, see \cite{Gao:2017yyd}.

Hard-scattering factorisation is central to our understanding of QCD, and is expressed through factorisation theorems that relate cross-sections (or experimental observables such as structure functions) to convolutions of a hard-scattering, perturbative kernel and nonperturbative PDFs. For example, the differential cross-section for unpolarised DIS can be written in terms of the structure functions $F_1$ and $F_2$ as 
\begin{equation}
\frac{\mathrm{d}^2\sigma}{\mathrm{d}x\,\mathrm{d}Q^2} = \frac{4\pi \alpha^2}{xQ^4}\left[\left(1-\frac{Q^2}{xs}-\frac{Q^2M^2}{s^2}\right)F_2(x,Q^2) - y^2F_1(x,Q^2)\right],
\end{equation}
up to higher order corrections in the electron mass and the electromagnetic coupling $\alpha$. Here the invariant momentum transfer is $Q = \sqrt{-q^2}$, $s$ is the centre-of-mass energy of the collision, and $x = Q^2/(2P\cdot q)$, with $P$ the momentum of the initial state hadron of mass $M$. The factorisation formulae for the structure functions are then
\begin{align}
F_1(x,Q^2) = {} & \sum_{i=q,\overline{q},g} \int_{x-}^{1+}\frac{\mathrm{d}\xi}{\xi}c_1\left(\frac{x}{\xi},\frac{Q^2}{\mu^2},\alpha_s\right)f_{i/H}(\xi,\mu)+\ldots, \label{eq:F1}\\
F_2(x,Q^2) = {} & \sum_{i=q,\overline{q},g} \int_{x-}^{1+}\mathrm{d}\xi \,c_2\left(\frac{x}{\xi},\frac{Q^2}{\mu^2},\alpha_s\right)f_{i/H}(\xi,\mu)+\ldots, \label{eq:F2}
\end{align}
where the limits account for potential endpoint singularities \cite{Collins:2011zzd}. The $c_{1,2}$ are the perturbative hard-scattering coefficients, which depend on the scattering probes and the $f_{i/H}$ are the nonperturbative PDFs, which characterise the longitudinal momentum structure of the target hadron. 

Collinear factorisation is crucial here: by separating out the hard scattering contribution, the remaining PDFs are independent of the external probe and can be properly considered universal. PDFs can therefore be extracted from simultaneous fits to a wide array of experimental processes. These fits are carried out by multiple groups, using a variety of techniques (for reviews see, for example, \cite{Gao:2017yyd,Lin:2017snn}), and the agreement or disagreement between the results provides some measure of systematic uncertainties in our knowledge of PDFs. For those PDFs for which we have extensive experimental data, such as the $u(x,Q^2)$ PDF at moderate values of $x\sim 0.1-0.5$, different PDF sets generally have small uncertainties and are in good agreement, suggesting small systematic uncertainties. In contrast, for PDFs for which there is little experimental data, such as the $\overline{d}(x,Q^2)$ PDF, different determinations have large uncertainties and collectively show significantly more tension, indicating a greater systematic uncertainty from the fitting procedure.

In the absence of extensive experimental data, first principles' calculations of PDFs will significantly improve PDF uncertainties \cite{Gao:2017yyd}. The interplay between lattice QCD and global fits was studied in \cite{Lin:2017snn}, where it was demonstrated that the impact of lattice calculations of both the lowest Mellin moments (see Sec.~\ref{sec:many}) and the $x$-dependence of PDFs could significantly reduce uncertainties in global PDF fits. For example, lattice determinations of the $\overline{d}(x,Q^2)$ PDF at moderate values of $x$ with uncertainties of $5-10\%$ could reduce the corresponding PDF uncertainties by up to $30-50\%$. This is of particular significance for several important new physics search channels at the LHC \cite{Gao:2017yyd,Lin:2017snn}.

\section{\label{sec:lattPDFs}PDFs from lattice QCD}

Over the last five years, a variety of nonperturbative approaches to $x$-dependent hadron structure have been studied, based on ideas both new and old. Much of the work has focussed on the quasi and pseudo PDF methods, but factorisable matrix elements, direct calculations of the Euclidean hadronic tensor and the forward Compton amplitude, 
and techniques to extract multiple Mellin moment have also been investigated. 

Three broad themes have emerged. These themes are not distinct; they are multifaceted and intersect in multiple ways, but they provide one framework through which we can view the issues raised by recent work. The first theme is the challenge of systematic uncertainties, which are now beginning to be explored \cite{Liu:2018uuj,Cichy:2018int,Braun:2018brg,Bali:2018spj}. Uncertainties such as discretisation and finite volume effects are common to all lattice calculations, but may be exacerbated for hadron structure calculations, many of which rely on highly boosted nucleons or nonlocal operators, or both. Conceptual and theoretical challenges have been studied in depth for the the LaMET and pseudo PDF approaches, including the renormalisation of nonlocal operators, matching, and factorisation. Higher twist effects and inverse transform problems are common to most, if not all, of the approaches, and are simultaneously conceptual challenges, systematic uncertainties and point to the third theme, the future role of precision lattice calculations. Higher twist effects, for example, are both systematic uncertainties and objects of primary study, relevant far beyond the lattice, as are the problems associated with inverse transforms from limited data. But these difficulties suggest new ways to envisage the role of lattice calculations. 

In the long term, one can anticipate that computational and algorithmic advances will reduce systematic uncertainties to the few percent level. In this future precision era, the inverse problems may be the last obstacle to direct computations of PDFs. Incorporating lattice data for matrix elements, rather than PDFs, into global fits, perhaps by fitting lattice results to an inverse Fourier transform of the PDF fit function, may provide one method to overcome the challenge of the Fourier transform. But perhaps new frameworks that avoid inverse problems entirely will be used. For example, the long view may prove that Ioffe time matrix elements (discussed in the next subsection) themselves are the most relevant quantities to characterise nonperturbative hadron structure, rather than PDFs. 

The possibility of a precision era also indicates that providing data for global fits need not be the main motivation for lattice calculations. Most obviously, nonperturbative calculations of more complicated quantities, such as generalised parton distributions and transverse momentum-dependent distributions (TMDs), will be particularly important in the electron-ion collider era. Lattice QCD will also be able to directly calculate higher twist effects and Ioffe time matrix elements at unphysical pion masses or kinematics, both of which will help limn the limits of the collinear factorisation paradigm itself \cite{Braun:2018brg,Moffat:2017sha}. In short, a precision era will bring a wealth of opportunities for lattice calculations of hadron structure.

\subsection{\label{sec:matelem}Ioffe time matrix elements}

PDFs arise directly in the context of collinear factorisation, but there is, arguably, a more natural starting point for our discussion, the ``Ioffe time distributions'' \cite{Braun:1994jq}. These are the Lorentz invariant matrix elements\footnote{For brevity I concentrate here on isovector quark distributions, which do not mix under renormalisation. Similar considerations apply for gluon distributions and mixing can be incorporated straightforwardly.}
\begin{equation}
M_\mu^{(0)}(n,P) = \left\langle P\left|\overline{\psi}(n)W(n,0)\Gamma_\mu  \psi(0) \right|P\right\rangle = 2P_\mu h_\Gamma^{(0)}(\nu,n^2)+z_\mu \overline{h}_\Gamma^{(0)}(\nu,n^2),
\end{equation}
where Lorentz invariance dictates the second equality and the dependence on the Ioffe time $\nu = n\cdot P$ \cite{Braun:1994jq}, and Wilson line length $n^2$. The superscript $(0)$ denotes a bare matrix element, $n_\mu$ is a spacetime four vector, and $\Gamma_\mu$ is a combination of Dirac gamma matrices, dictated by the Lorentz structure of the corresponding distribution. The target state, $|P\rangle$, is an exact momentum eigenstate, spin-averaged for unpolarised matrix elements, and usually taken to have relativistic normalization. Finally, the Wilson-line operator is
\begin{equation}
W(n,0) = {\cal P}\exp\Big[-ig_0\int_0^n\mathrm{d}y^\alpha \,A_\alpha^a(y)T^a\Big],
\end{equation}
where ${\cal P}$ represents path-ordering, and $A_\alpha(y) = A_\alpha^a(y) T^a$ is an $SU(3)$ gauge field in the fundamental representation,
with summation over the colour index $a$ implicit. 

\begin{figure}
\caption{\label{fig:factorisation}(Left) Schematic representation of the four objects discussed in this section and their relationships. (Right) The approach to the light cone in LaMET.}
\includegraphics[width=0.49\textwidth]{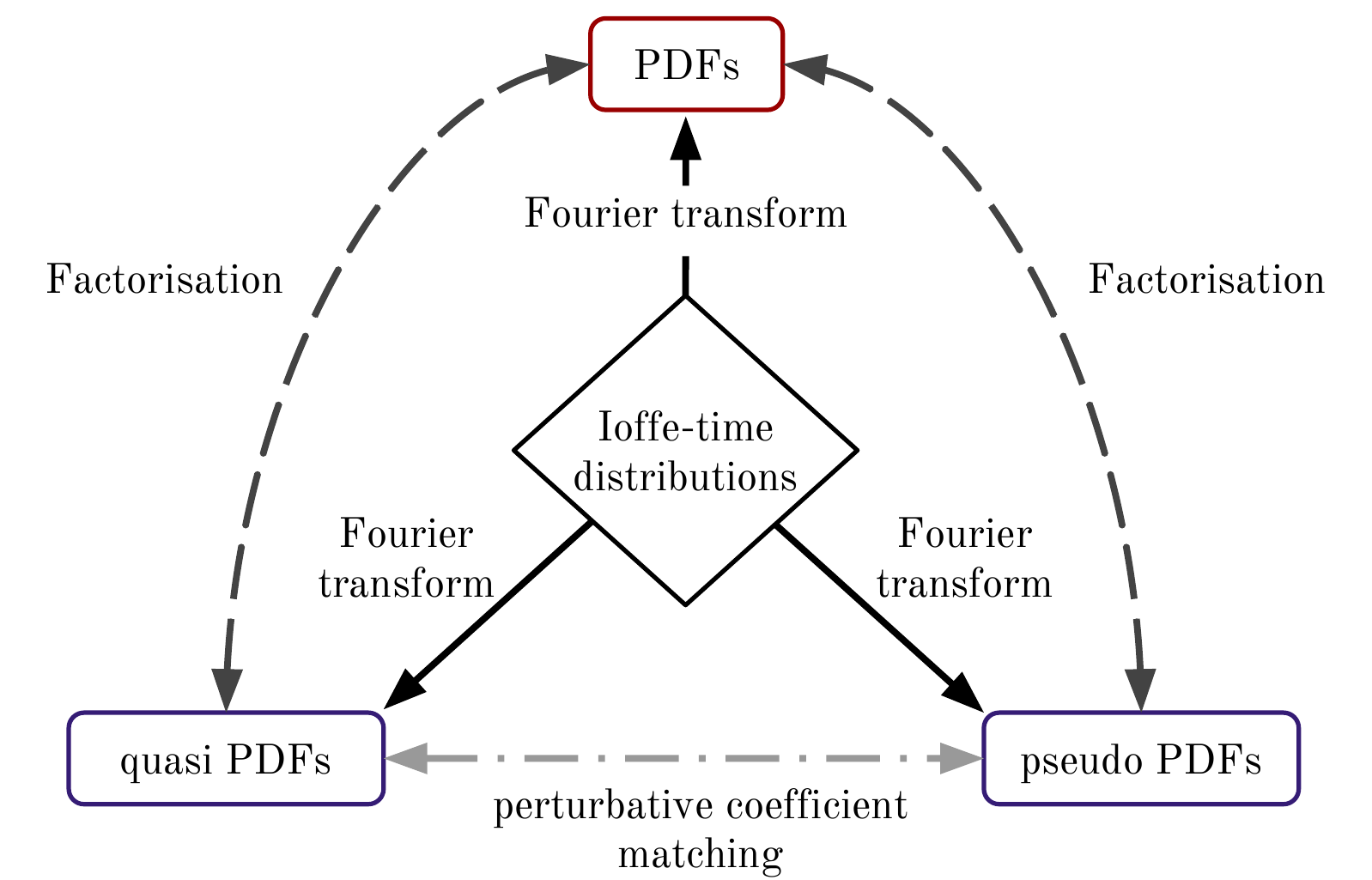}
\includegraphics[width=0.49\textwidth,keepaspectratio]{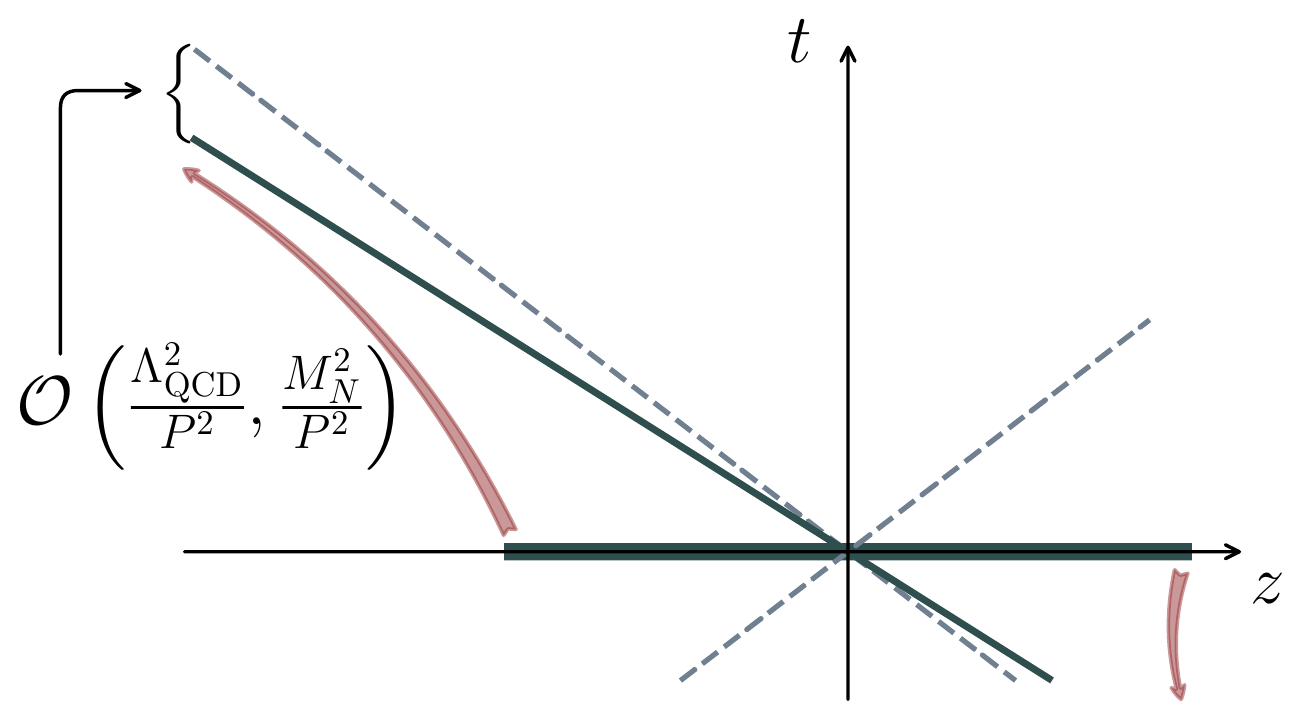}
\end{figure}

PDFs, and the closely related quasi PDFs \cite{Ji:2013dva} and pseudo PDFs \cite{Radyushkin:2017cyf} can be defined in terms of these Ioffe time matrix elements\footnote{In this review I will use the term ``Ioffe-time matrix elements'' to distinguish the matrix element from the distributions, which are defined by Fourier transform of the Ioffe-time matrix element and have an interpretation in terms of probability distributions (at least at leading order).} by particular choices of field separations $n^2$ and different Fourier transforms, represented schematically in the left hand panel of Fig.~\ref{fig:factorisation}.
Bare PDFs are defined by choosing $n^\mu = (0,n^-,\mathbf{0})$, so that $n^2=0$, and Fourier transforming with respect to $n^-$:
\begin{equation}
f^{(0)}_{i/H}(x) = \int_{-\infty}^\infty \frac{\mathrm{d}n^-}{4\pi}e^{-i\xi P^+n^-}\left\langle P\left|\overline{\psi}(0,n^-,\mathbf{0}_{\mathrm{T}})W(n^-,0)\Gamma \psi(0) \right|P\right\rangle = \int_{-\infty}^\infty \frac{\mathrm{d}\nu}{2\pi}e^{-i\xi \nu} h_\Gamma^{(0)}(\nu,0). \label{eq:pdfdef}
\end{equation}
Here I use light front coordinates $v^{\pm} = (v^0\pm v^3)/\sqrt{2}$ and $\mathbf{v}_{\mathrm{T}} = (v^1,v^2)$, with $v^\mu = (v^0,v^1,v^2,v^3)$.

In contrast, quasi  and pseudo PDFs are defined by choosing $n^\mu = (0,0,z,0)$, so that $n^2=-z^2$, and Fourier transforming with respect to the space-like separation $z$ and the Ioffe time $\nu$, respectively:
\begin{align}
q^{(0)}_{i/H}(\xi,P^z) = {} & \int_{-\infty}^\infty \frac{\mathrm{d}z}{4\pi}e^{i\xi P^zz}\left\langle P\left|\overline{\psi}(z)W(z,0)\Gamma \psi(0) \right|P\right\rangle  = E_{\mathbf{P}} \int_{-\infty}^\infty \frac{\mathrm{d}z}{2\pi}e^{i\xi P^zz}h_\Gamma^{(0)}(-P^zz,-z^2), \label{eq:qdef}\\ 
p^{(0)}_{i/H}(\xi,z^2) = {} & \int_{-\infty}^\infty \frac{\mathrm{d}\nu}{4\pi P^z}e^{-i\xi\nu }\left\langle P\left|\overline{\psi}(z)W(z,0)\Gamma \psi(0) \right|P\right\rangle = \int_{-\infty}^\infty \frac{\mathrm{d}\nu}{2\pi}e^{-i\xi\nu }h_\Gamma^{(0)}(\nu,-z^2). \label{eq:pdef}
\end{align}
By comparing Eqs.~\eqref{eq:pdfdef} and \eqref{eq:pdef}, we see that we can view pseudo PDFs as off the light cone generalisations of PDFs. Central to the quasi and pseudo PDF programmes is the fact that matrix elements of composite operators of spacelike-separated fields can be determined from an LSZ reduction in Minkowski spacetime, or from the long Euclidean time behaviour of Euclidean correlators \cite{Briceno:2017cpo}. Thus, lattice calculations of these matrix elements can be directly related to objects in Minkowski spacetime.

Fig.~\ref{fig:factorisation} represents not just the relationship between the Ioffe-time matrix element and the distributions, but also the relationships between the distributions themselves. These factorisation relations are central to the entire programme: factorisation enables the determination of light cone PDFs directly from matrix elements of space-like operators \cite{Ji:2017rah,Izubuchi:2018srq}. In particular, light cone PDFs can be extracted from renormalised quasi  and (reduced) pseudo PDFs via
\begin{align}
q_{i/H}(\xi,P^z,\mu_{\mathrm{R}}) = {} & \int_{-1}^1 \frac{\mathrm{d}y}{|y|}C^{(\widetilde{q})}\left(\frac{\xi}{y},\frac{\mu_{\mathrm{R}}}{\mu},\frac{\mu}{yP^z}\right)f_{j/H}(y,\mu)+{\cal O}\left(\frac{M^2}{(P^z)^2},\frac{\Lambda_{\mathrm{QCD}}^2}{(P^z)^2}\right) ,\label{eq:qfac}\\ 
p_{i/H}(\xi,z^2,\mu_{\mathrm{R}}) = {} & \int_{-1}^1 \frac{\mathrm{d}y}{|y|}C^{(\widetilde{p})}\left(\frac{\xi}{y},\frac{\mu_{\mathrm{R}}^2}{\mu^2},\mu^2 z^2\right)f_{j/H}(y,\mu)+{\cal O}\left(M^2z^2,\Lambda_{\mathrm{QCD}}^2z^2\right), \label{eq:pfac}
\end{align}
respectively. Both quasi and pseudo PDFs share the same collinear divergences as light cone PDFs, and this ensures that the factorisation coefficients, $C^{(\widetilde{q})}$ and $C^{(\widetilde{p})}$, are infrared safe and therefore can be determined in perturbation theory. Moreover, the coefficients can be related to each other through the perturbative relation
\begin{equation}
C^{(\tilde{f})}\left(\xi,\frac{\mu}{yP^z}\right) = \int \frac{\mathrm{d}\xi}{2\pi}e^{i\xi \zeta}\int_{-1}^1\mathrm{d}\alpha\,e^{-i\alpha \zeta}C^{(\tilde{p})}\left(\alpha,\frac{\mu^2\zeta^2}{(yP^z)^2}\right),
\end{equation}
represented in Fig.~\ref{fig:factorisation} by the dot-dashed grey horizontal double arrow. 

One can view these factorisation formulae, Eqs.~\eqref{eq:pdfdef} and \eqref{eq:pdef}, as arising because the $z^2\to 0$ limit of the Ioffe time matrix element is nontrivial; there is a logarithmic singularity. Quasi PDFs have support for $|x| > 1$, intimately tied to this perturbative $\ln(z^2)$ behaviour \cite{Radyushkin:2018nbf}, while PDFs and pseudo PDFs are restricted to $|x| \leq 1$ \cite{Radyushkin:2016hsy}. The behaviour of quasi PDFs in the unphysical $|x|>1$ region can be analysed through the relationship between quasi PDFs and TMDs \cite{Radyushkin:2017cyf,Musch:2010ka,Radyushkin:2018nbf} and it is argued in \cite{Radyushkin:2018nbf} that the $|x|>1$ region can be ignored altogether in lattice calculations of quasi PDFs. Inside the physical $x$ region, quasi and pseudo PDFs have different power corrections and thus may have greater precision in different ranges of $x$ \cite{Braun:2018brg}.

The factorisation formulae help clarify both the similarities and the differences between quasi and pseudo PDFs. Quasi PDFs are one example of the large momentum effective theory (LaMET) approach to light cone physics \cite{Ji:2014gla}, schematically represented in the right hand panel of Fig.~\ref{fig:factorisation}. PDFs, defined through composite operators of fields at light-like separations (represented by the $t = \pm z$ diagonal dashed grey lines), can be approximated by ``boosting'' spacelike operators, \emph{i.e.}~calculating matrix elements of spacelike operators between boosted states. The thick grey horizontal line represents the space-like extended operator, with boosting represented by the red curved arrows, and the resulting matrix element shown as the diagonal grey line. LaMET systematically accounts for the discrepancy between the resulting distributions at finite momentum (indicated by the ${\cal O}(\Lambda_{\mathrm{QCD}}/P^2,M_N^2/P^2)$ annotation) through the factorisation formula, Eq.~\eqref{eq:qfac}. In the pseudo PDF approach, in contrast, the light cone is approached in the limit $z^2\to 0$, at fixed Ioffe time $\nu$. 

It is worth pausing here to comment on the role of the hadron momentum. From a practical standpoint, the LaMET truncation uncertainty in the factorisation formulae is reduced by increasing the target state momentum. The signal-to-noise ratio in lattice calculations generally decays exponentially as $\exp[-(E(\mathbf{p})-3m_\pi/2)t]$ \cite{Parisi:1984ggi}, so that, even with the advent of momentum smearing \cite{Bali:2016lva}, high-momentum nucleons generally suffer from significant signal-to-noise difficulties in lattice calculations. Noisy lattice data at large nucleon momentum is an impediment to reducing the corresponding finite momentum systematic uncertainties in the factorisation formulae for quasi PDFs, although models suggest momenta of \SIrange{2}{3}{GeV} may be sufficient \cite{Gamberg:2014zwa}. For pseudo PDFs, on the other hand, the corrections scale as ${\cal O}(\Lambda_{\mathrm{QCD}}^2z^2)$, and reducing the corresponding systematic uncertainty requires balancing the length of the operator with the nucleon momentum. Thus, the momentum does not suppress power corrections, but provides a lever arm to access a wider range of Ioffe time at small $z^2$, required to suppress corrections but limited by the finite lattice spacing. Lastly, as we will see, the hadron momentum plays varying roles in the other approaches. For factorisable matrix elements, it plays a similar role to that for pseudo PDFs, while for the hadronic tensor and Compton amplitude methods, large momenta are required to reach the DIS region.

\subsubsection{\label{ssec:renorm}Renormalisation}

The continuum Wilson line operator that appears in the Ioffe time matrix element is multiplicatively renormalisable in position space \cite{Ishikawa:2017faj,Zhang:2018diq}. The lattice regulator introduces mixing, first pointed out in \cite{Constantinou:2017sej} and studied in more detail in \cite{Chen:2017mie}, but I focus on continuum operators. This point, along with the factorisation formulae, Eqs.~\eqref{eq:qfac} and \eqref{eq:pfac}, is one of the central insights understood over the last year. Early work on the renormalisation of quasi PDFs conjectured a convolution relation for the renormalised matrix element in momentum space, which is now understood to be incorrect\footnote{The work of \cite{Rossi:2018zkn} can be understood as a proof by contradiction, starting from the convolution Ansatz, although some of the conclusions reached in that work are incorrect \cite{Ji:2017rah,Radyushkin:2018nbf,Karpie:2018zaz}.}. Indeed, it is important to distinguish three separate conceptual issues that were somewhat conflated in early work (in practice matching and factorisation may be carried out as a single step): the nonperturbative renormalisation of the nonlocal operator (\emph{i.e.}~removing power and logarithmic divergences via, for example, the RI-MOM scheme) \cite{Radyushkin:2017cyf,Musch:2010ka,Ishikawa:2017faj,Zhang:2018diq,Xiong:2017jtn,Ji:2015jwa,Monahan:2016bvm,Alexandrou:2017huk,Spanoudes:2018zya}; scheme matching of renormalised matrix elements or distributions (\emph{i.e.}~converting, say, the quasi PDF in the RI/MOM scheme to the $\ms$ scheme) \cite{Liu:2018tox,Stewart:2017tvs,Alexandrou:2018pbm}; and factorisation of renormalised distributions (\emph{i.e.}~extracting renormalised PDFs from renormalised quasi PDFs) \cite{Ji:2017rah,Izubuchi:2018srq,Monahan:2016bvm,Stewart:2017tvs,Xiong:2013bka,Ma:2014jla}. These issues have each received considerable attention. I highlighted factorisation in the previous section and now turn to renormalisation. Scheme matching introduces technical issues \cite{Stewart:2017tvs,Alexandrou:2018pbm} beyond the scope of this review.

The bare Wilson-line operator has both logarithmic and power divergences, where the latter is associated with the length of extended Wilson line \cite{Ishikawa:2017faj,Zhang:2018diq}. The power divergence must be removed nonperturbatively to define the continuum limit of lattice-regulated matrix elements. There have been various approaches suggested, which I classify into four broad categories: 1)~exponentiated mass counterterm; 2)~RI schemes; 3)~the ratio method; and 4)~the gradient flow.

\paragraph{Exponentiated mass counterterm} The close relationship between Wilson lines and static heavy quarks, and heavy quark effective theory (HQET), has informed studies of various aspects of the nonlocal operator, $O_\Gamma = \overline{\psi}(n)W(n,0)\Gamma  \psi(0) $.

In particular, based on early work on Wilson lines \cite{Dotsenko:1979wb}, the authors of Ref.~\cite{Musch:2010ka,Ishikawa:2016znu} conjectured that the nonlocal bilinear operator, $O_\Gamma$ renormalises according to $O_\Gamma = Z_\psi^{-1} Z_z^{-1} e^{-\delta m L(C)}O_\Gamma^{(0)}$, where the $Z_\psi$ is the quark wavefunction renormalisation, $\delta m$ is the mass renormalisation of a test particle moving along the smooth contour $C$ of length $L$ that defines the Wilson line path, and $Z_z$ is a logarithmic renormalisation parameter associated with the endpoints of the Wilson line. The exponentiated mass counterterm $\delta m$ corresponds to the additive heavy quark mass renormalisation in HQET and removes all power divergences in the bare operator $O_\Gamma^{(0)}$. Based on this renormalisation pattern, one can define a ``power divergence subtracted'' matrix element $h_\Gamma^{\mathrm{PS}}(\nu,n^2) = e^{-\delta m |n|}h_\Gamma(\nu,n^2)$. The mass counterterm can be determined from the static heavy quark potential \cite{Ishikawa:2016znu} or through ratios of bare matrix elements \cite{Constantinou:2017sej}.

\paragraph{RI schemes} Regularisation-independent (RI) schemes have a long pedigree as a nonperturbative renormalisation scheme for local operators \cite{Martinelli:1994ty}. Last year, RI schemes were proposed for nonlocal bilinear operators \cite{Alexandrou:2017huk,Spanoudes:2018zya}. These schemes offer a number of advantages. First, matching to the $\ms$ scheme (the scheme in which  phenomenologically-determined PDFs are traditionally expressed) can be done in the continuum, which avoids the need for lattice perturbation theory. Second, these schemes nonperturbatively remove both power and logarithmic divergences. There are, however, two challenges associated with RI schemes: they break gauge invariance, because they are defined via matrix elements with external off-shell quark or gluon states, and they introduce new discretisation effects through the external state momentum.

For operators that do not mix, the operator renormalisation parameters in the RI$^\prime$ scheme, $Z_{O_\Gamma}(z)$, are defined through (analogous relations hold for the RI/MOM and RI schemes)
\begin{equation}
Z_\psi^{-1} Z_{O_\Gamma}(z)\frac{1}{12}\mathrm{Tr}\,\left[\frac{{\cal V}(p,z)}{{\cal V}^{\mathrm{tree}}(p,z)}\right]_{p^2=\overline{\mu}^2} = 1,\qquad Z_\psi = \frac{1}{12}\mathrm{Tr}\,\left[\frac{S^{\mathrm{tree}}(p)}{S(p)}\right]_{p^2=\overline{\mu}^2}.
\end{equation}
In these equations the trace is taken over both spin and colour indices and the momentum $p$ defines the RI$^\prime$ renormalisation scale $\overline{\mu}$. The amputated vertex function is ${\cal V}(p,z)$, the quark propagator is $S(p)$, and the superscript ``tree'' indicates the corresponding tree-level values. There are several options for the choice of the momentum, $p$, and the RI$^\prime$ scale, $\overline{\mu}$, which are discussed in more detail in \cite{Alexandrou:2017huk}, with certain choices generally suppressing or enhancing discretisation effects.

\paragraph{Ratio method} The ratio method, or ``reduced pseudo PDF'' method was introduced in \cite{Radyushkin:2017cyf}. The power divergence is removed from the Ioffe-time matrix element by dividing by the same matrix element, evaluated at zero Ioffe time, $h_\Gamma^{\mathrm{red}}(\nu,z^2) = h_\Gamma(\nu,z^2)/h_\Gamma(0,z^2)$. This procedure removes both power and logarithmic ultraviolet (UV) divergences, so that the continuum limit of the reduced matrix element is well-defined and largely scale independent, up to a vestigial logarithmic $z^2$ dependence that corresponds to the DGLAP evolution of PDFs \cite{Radyushkin:2017cyf}. Preliminary nonperturbative results corroborate these perturbative expectations \cite{Orginos:2017kos}, and are illustrated in Fig.~\ref{fig:ppdf}.

\paragraph{Gradient flow} The gradient flow is a one parameter mapping that reduces UV field fluctuations, corresponding to smearing in real space \cite{Narayanan:2006rf}. The mapping parameter is generally called the flow time, and the corresponding smearing radius scales with the flow time, $\tau$, as $\sqrt{2D\tau}$ in $D$ dimensions. Smearing has a long history in lattice calculations, with a concomitant wide variety of applications, but the gradient flow provides a particularly useful property: guaranteed finiteness. Up to a multiplicative fermionic wavefunction renormalisation, renormalised correlation functions are guaranteed to remain finite at finite flow time \cite{Luscher:2011bx}.

The gradient flow was proposed as a mechanism to remove the power divergence of the nonlocal bilinear operator in \cite{Monahan:2016bvm}. The matrix element of the smeared nonlocal operator $O_\Gamma$ is finite and has a well-defined continuum limit. Then, provided $\Lambda_{QCD},M_N\ll P_z \ll \tau^{-1/2}$, PDFs can be factorised from the the continuum smeared distribution, using a formula analogous to Eqs.~\eqref{eq:qfac} and \eqref{eq:pfac}, because the flow time serves as a UV regulator and does not affect the IR behaviour of the distributions. Based on the exponential behaviour of the Wilson line outlined above, it is expected that the power divergence can be determined by a nonperturbative fit of the flow time dependence at fixed $z^2$. This has yet to be demonstrated in practice.

\subsubsection{\label{ssec:qpdferrors} Systematic uncertainties}
\begin{figure}
\centering
\caption{\label{fig:flowchart}Schematic representation of the work flow for determining PDFs (red rounded box on the right) directly from lattice calculations of Ioffe time matrix elements (dark blue rounded box on the left). The intermediate objects are represented the rounded light purple boxes, moving from left to right, with the steps required highlighted by the grey boxes at the top (and labelled by the dashed arrows). The corresponding systematic uncertainties appear in the rounded boxes (and dot-dashed arrows) at the bottom. For more details of each step and systematic uncertainty, see the accompanying text.}
\includegraphics[width=\textwidth,keepaspectratio]{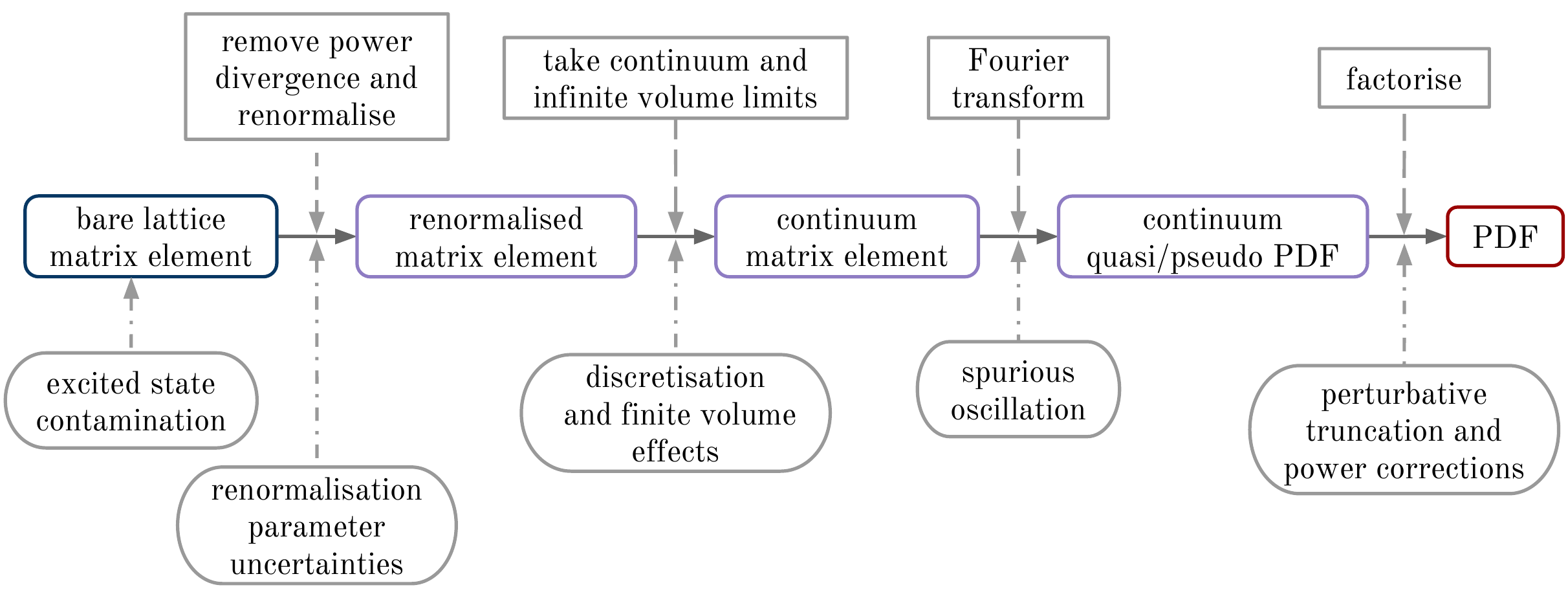}
\end{figure}

Lattice calculations are necessarily carried out at finite lattice spacing in finite volumes and there are corresponding systematic uncertainties in all lattice calculations. Nonperturbative calculations of quasi and pseudo PDFs in particular, however, may suffer from enhanced excited state effects, discretisation errors and finite volume effects, in addition to the more studied perturbative truncation and finite momentum corrections. The extraction of PDFs from lattice calculations of Ioffe time matrix elements involves a succession of steps, schematically represented in Fig.~\ref{fig:flowchart}, each of which can be viewed as introducing associated systematic uncertainties. As lattice calculations of $x$-dependent hadron structure move towards maturity, each of the following sources of systematic uncertainty must be studied, quantified and controlled.

\paragraph{Excited state contamination} Nonperturbative matrix elements are extracted from the long Euclidean time behaviour of Euclidean correlators. Long Euclidean times are saturated by the ground state, but at short and intermediate Euclidean times, the correlation functions have excited state contributions. In spectroscopic studies, excited states can be disentangled using, for example, a large basis of interpolating operators \cite{Dudek:2009qf}, but in general $x$-dependent structure calculations are focussed on the extraction of the ground state. Unfortunately, baryons in particular suffer from significant signal-to-noise problems at large Euclidean times, which is exacerbated for boosted systems. This leads to challenging excited systematics in the extraction of ground state signals at intermediate Euclidean times. The traditional two-state fit approach to three-point functions appears to require nucleon source-sink separations of $\gtrsim\SI{1}{fm}$ for nucleon momenta around $P^z\simeq $\SIrange{1}{1.5}{GeV} \cite{Cichy:2018int}. The summation method offers an alternative approach, used, for example, in \cite{Orginos:2017kos}. The size of systematic uncertainties associated with excited state effects in Ioffe time matrix element calculations is still under investigation.

\paragraph{Discretisation effects} Potentially large discretisation effects are expected to arise from ${\cal O}(ap)^n$ contributions generated by the large nucleon momentum, and, in the case of RI schemes, by the RI renormalisation parameter. Most nonperturbative results currently available are at a single lattice spacing, and discretisation effects are generally unknown, although these have been studied to some extent in the factorisable matrix element approach \cite{Bali:2018spj} and for lattice TMD calculations \cite{Musch:2010ka}.

\paragraph{Finite volume effects} Enhanced finite volume effects may be generated by nonlocal operators \cite{Briceno:2018lfj}. In contrast to local operators, which generically have exponentially suppressed finite volume effects of the form $e^{- m_\pi L}$, there are two IR scales present, the lattice size $L$ and the spatial extent of the Ioffe-time operator, $\xi$. It is therefore natural to expect that the relevant length scale is not the lattice size itself, but the difference between the lattice size and operator extent. This was confirmed for spatially separated currents in a scalar model in \cite{Briceno:2018lfj}, where the leading order finite volume effects for the scalar analogue of the pion were found to scale as $e^{-m|L-\xi|}$. Although this work has to be generalised to operators that include Wilson lines, this suggests that these effects could be a significant source of systematic uncertainty on currently available lattices.

\paragraph{Spurious oscillations} Even with sufficient computing resources and algorithmic improvement to render discretisation, finite volume, and finite momenta effects negligible, lattice calculations will always be restricted to a finite number of determinations of the Ioffe time matrix element over a restricted range of Wilson line lengths. Perhaps the most challenging systematic uncertainty for this entire approach, then, is the spurious oscillations introduced by the Fourier transform of the matrix element. A number of approaches have been proposed to overcome this challenge, including a low-pass filter, the ``derivative method'', and Gaussian weighting for quasi PDFs \cite{Lin:2017ani} (see also \cite{Musch:2010ka}) and Backus-Gilbert, Bayesian and neural network reconstruction for pseudo PDFs \cite{Zafeiropoulos:2018lat}. It is not yet clear, however, if these oscillations can be completely controlled in a model independent way \cite{Cichy:2018int}. In some respects, the Fourier transform may represent a fundamental limitation of the current approach and perhaps suggests reframing how lattice calculations are incorporated into, or compared with, phenomenological extractions of PDFs.

\paragraph{Truncation effects and power corrections} Finally, perturbative truncation effects, especially at scales characteristic of current lattice calculations, around \SI{2}{GeV}, could be significant, since current matching and factorisation calculations include only one loop effects, although a two loop calculation appears in \cite{Ji:2015jwa}. Reducing this uncertainty will require two loop matching and factorisation calculations. Moreover, the nucleon momentum is generally limited to $\sim$\SI{2}{GeV}, so that power-suppressed corrections of the form $(\Lambda_{\mathrm{QCD}}/P^z)^2$ ($\Lambda_{\mathrm{QCD}}z^2$ in the pseudo PDF approach) in the factorisation formulae, may also be significant. The so-called target mass corrections, of ${\cal O}(M_N/P^z)^2$, can be removed analytically \cite{Chen:2016utp}. Truncation uncertainties and power corrections are perhaps the most studied of the systematics in the quasi PDF approach \cite{Liu:2018uuj,Cichy:2018int,Braun:2018brg}.

\subsubsection{\label{ssec:qpdfres} Recent results} 

\paragraph{Quasi PDFs} The first nonperturbative calculations were published in \cite{Lin:2014zya}, with a number of updates since \cite{Alexandrou:2017huk,Chen:2016utp,Alexandrou:2016jqi}, but this year marks a particularly important achievement: the first calculations of Ioffe time matrix elements at physical pion masses, from ETM \cite{Alexandrou:2018pbm} and LP$^3$ \cite{Lin:2017ani,Chen:2018xof,Lin:2018qky} collaborations. Although this is a significant advance, when viewed through the lens of the previous section, it is clear that a number of systematic uncertainties are yet to be addressed (some systematics are studied in \cite{Liu:2018uuj,Cichy:2018int}). In particular, most lattice calculations are at a single lattice spacing, volume, and pion mass.

\begin{figure}
\centering
\caption{\label{fig:qpdf_etmc}Isovector unpolarised (left) and transversity (right) distributions from \cite{Cichy:2018int,Alexandrou:2018pbm,Alexandrou:2018eet}.} 
\includegraphics[width=0.49\textwidth,keepaspectratio]{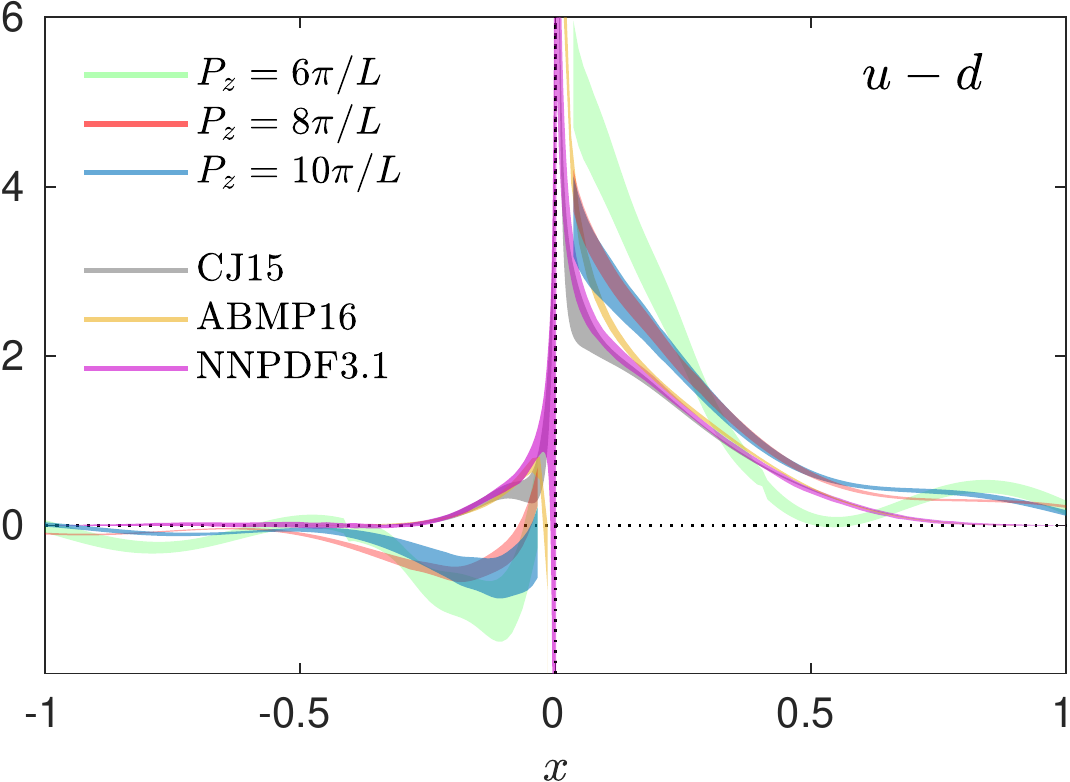}
\includegraphics[width=0.49\textwidth,keepaspectratio]{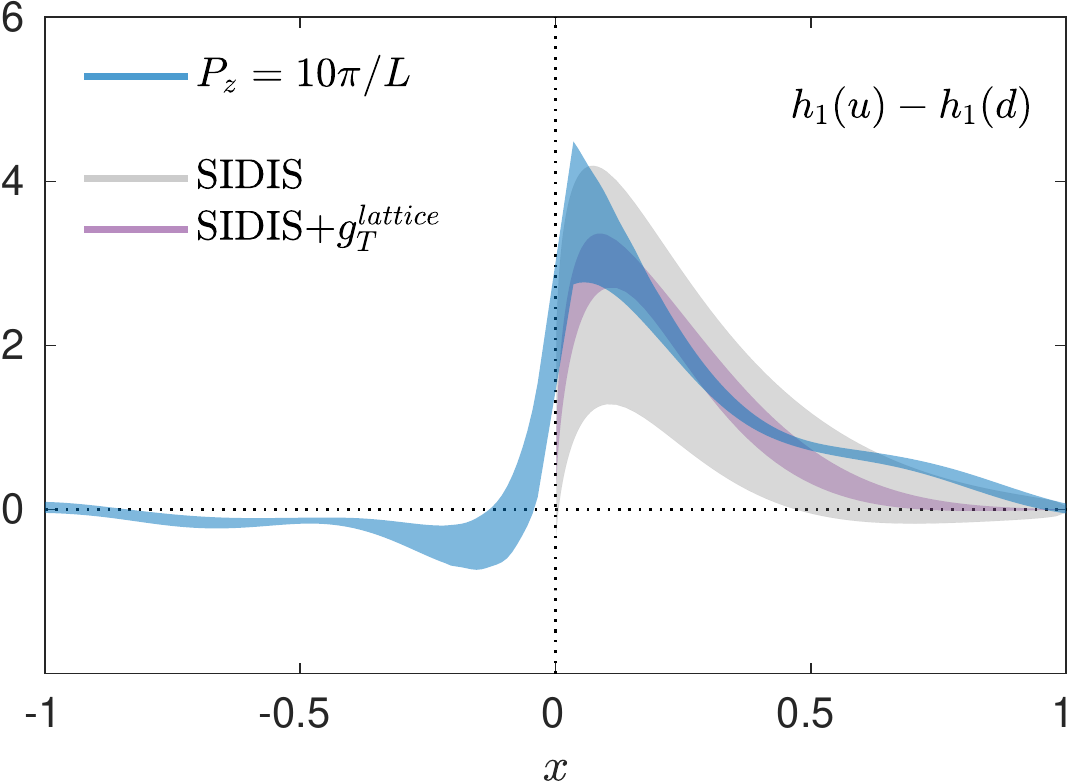}
\end{figure}

\begin{figure}
\centering
\caption{\label{fig:qpdf_lp3}Isovector unpolarised (left) and transversity (right) distributions from \cite{Chen:2018xof,Liu:2018hxv}.} 
\includegraphics[width=0.49\textwidth,keepaspectratio]{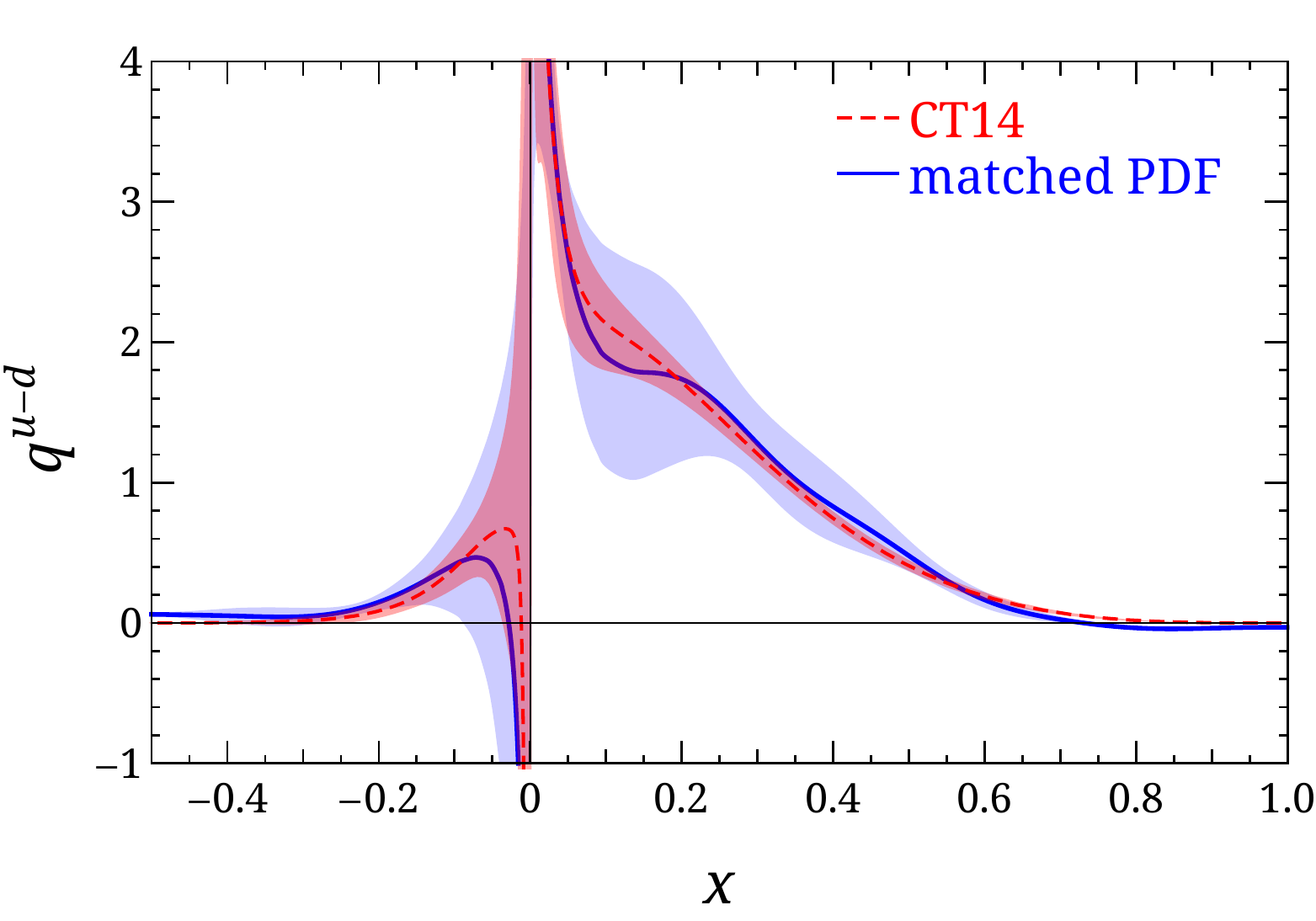}
\includegraphics[width=0.49\textwidth,keepaspectratio]{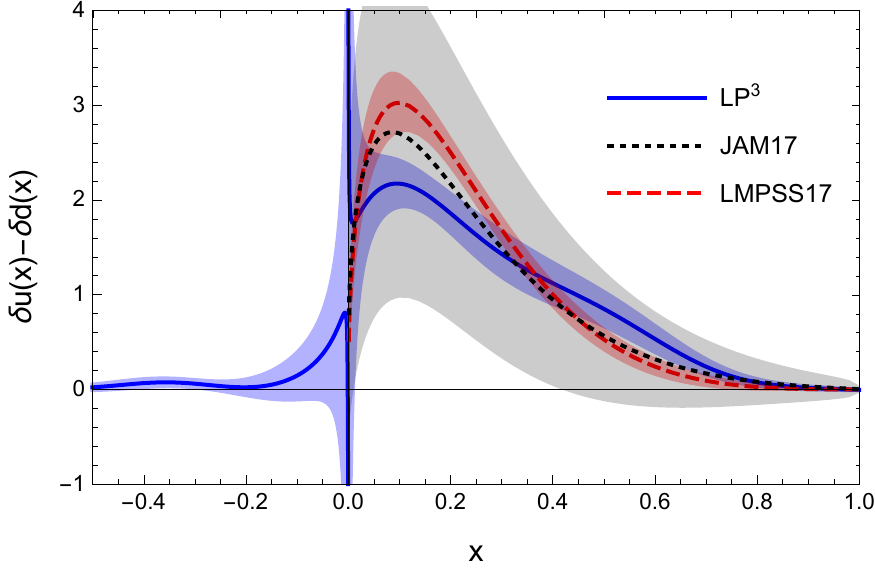}
\end{figure}

I illustrate the recent quasi PDF results at physical pion masses in Figs.~\ref{fig:qpdf_etmc} and \ref{fig:qpdf_lp3}. Fig.~\ref{fig:qpdf_etmc} shows the latest results for the isovector unpolarised  (left) and transversity (right) quark distributions from the ETM collaboration \cite{Cichy:2018int}, updated from \cite{Alexandrou:2018pbm,Alexandrou:2018eet}, computed on a single $n_f = 2$ twisted mass ensemble, with lattice spacing $a\simeq \SI{0.094}{fm}$, volume $L^3/a^3\times T/a = 48^3\times 96$, and pion mass $m_\pi = \SI{130.4(4)}{MeV}$. The left hand figure shows the isovector unpolarised quark PDF at three different lattice momenta, compared to three different phenomenological PDF fits. These momenta are $6\pi/L\simeq \SI{0.83}{GeV}$ (green band), $8\pi/L\simeq \SI{1.11}{GeV}$ (orange), and $10\pi/L\simeq \SI{1.38}{GeV}$ (blue), and it is clear that, although the error band does not include a complete error budget, at least qualitatively the bands move towards the phenomenological results as the nucleon momentum increases. The right hand plot presents the quark transversity PDF at a single lattice momentum $10\pi/L\simeq \SI{1.38}{GeV}$ as a light blue band, representing statistical uncertainties only. The figure also shows the corresponding transversity distribution extracted from phenomenological fits, the grey band labelled ``SIDIS'' (semi-inclusive DIS), and the transversity distribution extracted from experimental data and constrained by lattice calculations of the first moment, the tensor charge, labelled ``SIDIS+$g_{\mathrm{T}}^{\mathrm{lattice}}$''. The first moment determined by integrating the resulting PDF over $x$ is $g_{\mathrm{T}}^{\mathrm{qPDF}} = 1.09(11)$, in good agreement with the recent lattice calculations of this quantity, $g_{\mathrm{T}} = 1.09(11)$ and $g_{\mathrm{T}} = 0.989(32)(10)$ \cite{Alexandrou:2017qyt}.

In Fig.~\ref{fig:qpdf_lp3} I present the same distributions (isovector unpolarised and transversity distributions) calculated by the LP$^3$ collaboration \cite{Chen:2018xof,Liu:2018hxv}. The left hand figure shows the final, matched PDF result from a lattice calculation on an $n_f = 2+1+1$ MILC HISQ ensemble with $a\simeq \SI{0.09}{fm}$, $L^3/a^2\times T/a = 64^3\times 96$, and $m_\pi \simeq \SI{135}{MeV}$. The uncertainty band shown in blue includes an estimate of the finite momentum uncertainties, based on three lattice momenta $20\pi/L\simeq \SI{2.2}{GeV}$, $24\pi/L\simeq \SI{2.6}{GeV}$, and $28\pi/L\simeq \SI{3.0}{GeV}$. The right hand figure shows the isovector quark transversity distribution at $\mu^2_{\mathrm{R}}=\SI{2}{GeV}$, extracted from the largest lattice momentum of $P^z = \SI{3.0}{GeV}$, compared with two phenomenological fits. The blue error band includes both statistical and systematic uncertainties.

Both the ETMC and LP$^3$ results presented in Figs.~\ref{fig:qpdf_etmc} and \ref{fig:qpdf_lp3} incorporate nonperturbative renormalisation via RI$^\prime$ and RI/MOM schemes, respectively, with an operator Lorentz structure chosen to remove operator mixing at finite lattice spacing \cite{Constantinou:2017sej}. The LP$^3$ results use the ``derivative'' method \cite{Lin:2017ani} to reduce spurious oscillations in the Fourier transform. It is worth remarking that the ETMC results use a rather small lattice volume, with $m_\pi L \sim 3.0$, and, although LP$^3$ use a larger volume, with $m_\pi L \sim 4.0$, they use Wilson line operators of length $z_{\mathrm{max}}/a = 20$, so that the IR length scale $m_\pi|L-z_{\mathrm{max}}| \sim 2.7$ is quite small. Matching (to the $\ms$ scheme) and factorisation coefficients are known only to one loop in perturbation theory, leaving ${\cal O}(\alpha_S^2)$ truncation errors in all results. Thus, although the results of Figs.~\ref{fig:qpdf_etmc} and \ref{fig:qpdf_lp3} are qualitatively very encouraging, particularly for the transversity distribution, the uncertainty bands should be treated with some caution and proper quantitative comparison to phenomenological fits requires more data.

The focus of much of the numerical work on quasi PDFs has been on the isovector quark distributions, which are the simplest to compute, but gluon distributions have also been studied \cite{Zhang:2018diq,Wang:2017qyg}. Moving beyond nucleons, both distribution amplitudes (DAs) and PDFs of mesons have been calculated on the lattice, including for the pion \cite{Zhang:2017bzy,Chen:2018fwa} (see Fig.~\ref{fig:piDA}) and for the kaon \cite{Chen:2017gck}.

\begin{figure}
\centering
\caption{\label{fig:ppdf}(Left) The reduced Ioffe time matrix element from \cite{Zafeiropoulos:2018lat}. (Right) The heavy-light hadronic tensor from \cite{Detmold:2018kwu}.} 
\includegraphics[width=0.47\textwidth,keepaspectratio]{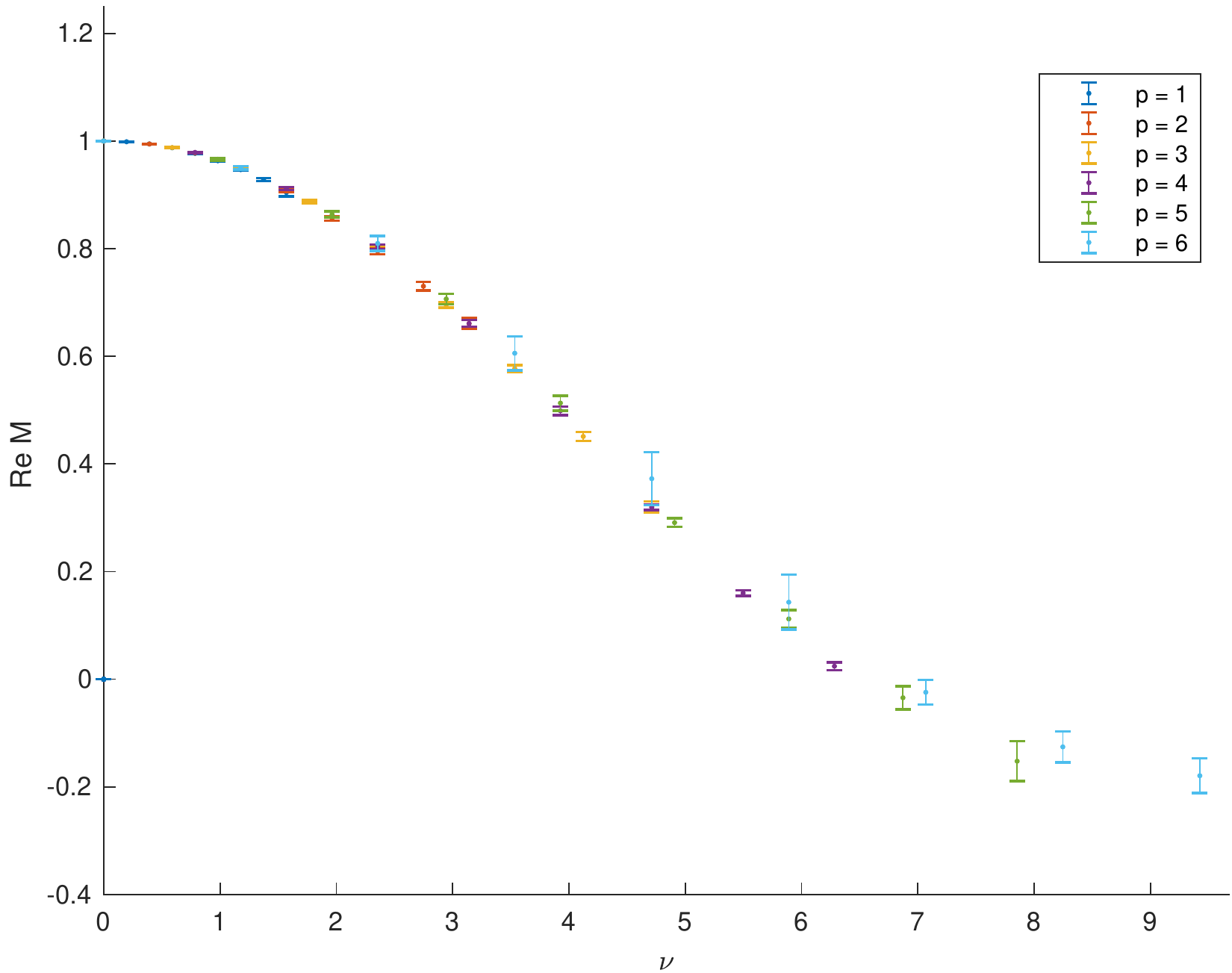}
\includegraphics[width=0.49\textwidth,keepaspectratio]{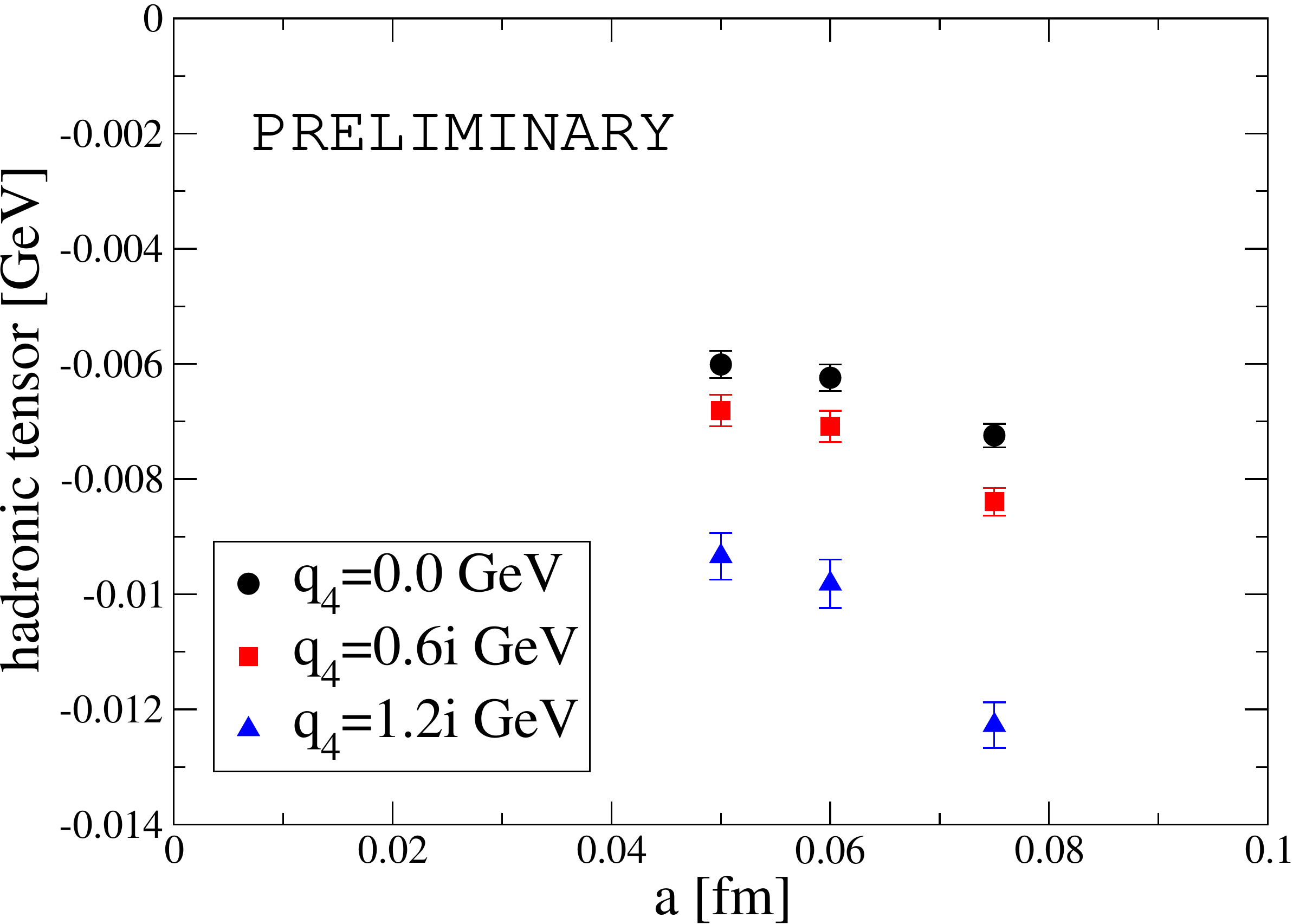}
\end{figure}

\paragraph{Pseudo PDFs} The first unquenched results for pseudo PDFs appear in \cite{Zafeiropoulos:2018lat}, which I show in the left hand plot in Fig.~\ref{fig:ppdf}. The first quenched results were presented in \cite{Orginos:2017kos}. The figure shows the Ioffe time matrix element at six values of the lattice momentum (in units of $2\pi/L$), as a function of the Ioffe time $\nu$,  from a lattice calculation on an $n_f = 2+1$ clover fermion ensemble with $a\simeq \SI{0.097}{fm}$, $L^3/a^3\times T/a = 32^3\times 64$, and $m_\pi \simeq \SI{450}{MeV}$. The data points include only statistical uncertainties, and the spread of points at fixed Ioffe time indicates the logarithmic scaling in $z^2$ of the matrix element \cite{Radyushkin:2017cyf}.

\subsection{\label{sec:facmat}Factorisable matrix elements}

Ioffe time matrix elements are an example of a more general approach to extracting PDFs from lattice calculations \cite{Ma:2014jla}, which I call ``factorisable matrix elements''\footnote{These objects have been referred to as ``good lattice cross-sections'', by analogy with the global fitting procedure, but these  quantities are not cross-sections. Therefore I will use the term factorisable matrix element, which is both technically correct and emphasises the key property of factorisation.}. In essence, one defines a set of matrix elements that are ``good'' in the sense that they: 1)~are calculable from Euclidean correlation functions; 2)~have a well-defined continuum limit; and 3)~share the same logarithmic collinear divergences as PDFs. Such matrix elements $\sigma_H$ then satisfy a general factorisation relation
\begin{equation}
\sigma_H(\nu,n^2,\mu_{\mathrm{R}}^2) = \sum_{i=q,\overline{q},g} \int_0^1 \frac{\mathrm{d}\xi}{\xi}C_i\left(\xi \nu, n^2,\mu_{\mathrm{R}}^2,\mu_{\mathrm{F}}^2\right)f_{i/H}(\xi,\mu_{\mathrm{F}}^2)+{\cal O}\left(\Lambda_{\mathrm{QCD}}^2n^2\right)
\end{equation}
where $\mu_{\mathrm{R}}$ and $\mu_{\mathrm{F}}$ are the renormalisation and factorisation scales, respectively. It is the third property, in particular, that ensures the coefficients $C_i$ are IR safe and perturbatively calculable. The factorisation formulae for the quasi and pseudo PDFs, Eqs.~\eqref{eq:qfac} and \eqref{eq:pfac}, are primary examples of this more general framework. 

Given a set of suitably chosen factorisable matrix elements $\sigma_H$, and a set of corresponding known perturbative factorisation coefficients $C_i$, PDFs can be extracted from a global fit to lattice data, in much the same way that PDFs are determined from experimental data.

This general approach was foreshadowed in \cite{Braun:2007wv}, which proposed extracting $x$-dependent hadron structure directly from position space correlators. For example, the pion DA, defined through
\begin{equation}\label{eq:phipi}
\phi_\pi(x) = \frac{i}{f_\pi}\int \frac{\mathrm{d}y}{2\pi} e^{i(x-1)\omega^-P^+}\left\langle | \overline{\psi}( 0,\omega^-,\mathbf{0}_{\mathrm{T}})W(\omega^-,0)\gamma^-\gamma_5 \psi(0) | \pi(P)\right\rangle,
\end{equation}
and normalised by the pion decay constant $f_\pi \simeq \SI{130}{MeV}$ such that $\int_0^1\mathrm{d}x\,\phi_\pi(x)= 1$, 
can be extracted from the equal-time matrix element \cite{Bali:2017gfr}
\begin{equation}
T(\nu,z^2) = \langle 0| J_5\left(-\frac{z}{2}\right) J_S\left(\frac{z}{2}\right)|\pi(P) \rangle = \frac{f_\pi \nu}{2\pi^2(z^2)^2}\int_0^1 \mathrm{d}x\,e^{i(x-1/2)\nu}C_\pi(x,z^2,\mu_{\mathrm{F}}^2)\phi_\pi(x,\mu_{\mathrm{F}}^2).
\end{equation}
Here $J_s = \overline{u}q$ and $J_5 = \overline{q}\gamma_5 u$, with $q\neq u,d$ a fictitious light quark flavour. The second equality assumes sufficiently small (spacelike) $z^2$, so that this correlation function can be determined in continuum perturbation theory, up to higher twist corrections, with $C_\pi = 1+{\cal O}(\alpha_S)$ a short-distance coefficient function and $\mu_{\mathrm{F}}$ the factorisation scale.

This approach has a number of potential advantages, perhaps the most important of which is that there is no Fourier transform to momentum space, although the Mellin integral (or its inverse) in Eq.~\eqref{eq:phipi} must be invoked to compare phenomenological data or model expectations to lattice results. Moreover, the bi-local current operator requires only local renormalisation parameters, there is no power divergence. The use of two currents may also reduce discretisation effects, by allowing off-axis separations to be used, which is a challenge for Wilson line operators, because of the presence of cusp anomalous dimensions. In common with the pseudo PDF approach, the smallness of higher twist contributions is controlled by small values of the current separation $z^2$. 

\subsection{\label{sec:facmatres}Results}

\begin{figure}
\centering
\caption{\label{fig:piDA} (Left) Pion DA parameterisations from \cite{Bali:2018spj}, with a comparison to the results of \cite{Zhang:2017bzy}. (Right) Preliminary results for the factorisable matrix element of two vector currents from \cite{Sufian:2018jla}. See the accompanying text for more details.}
\includegraphics[width=0.49\textwidth,keepaspectratio]{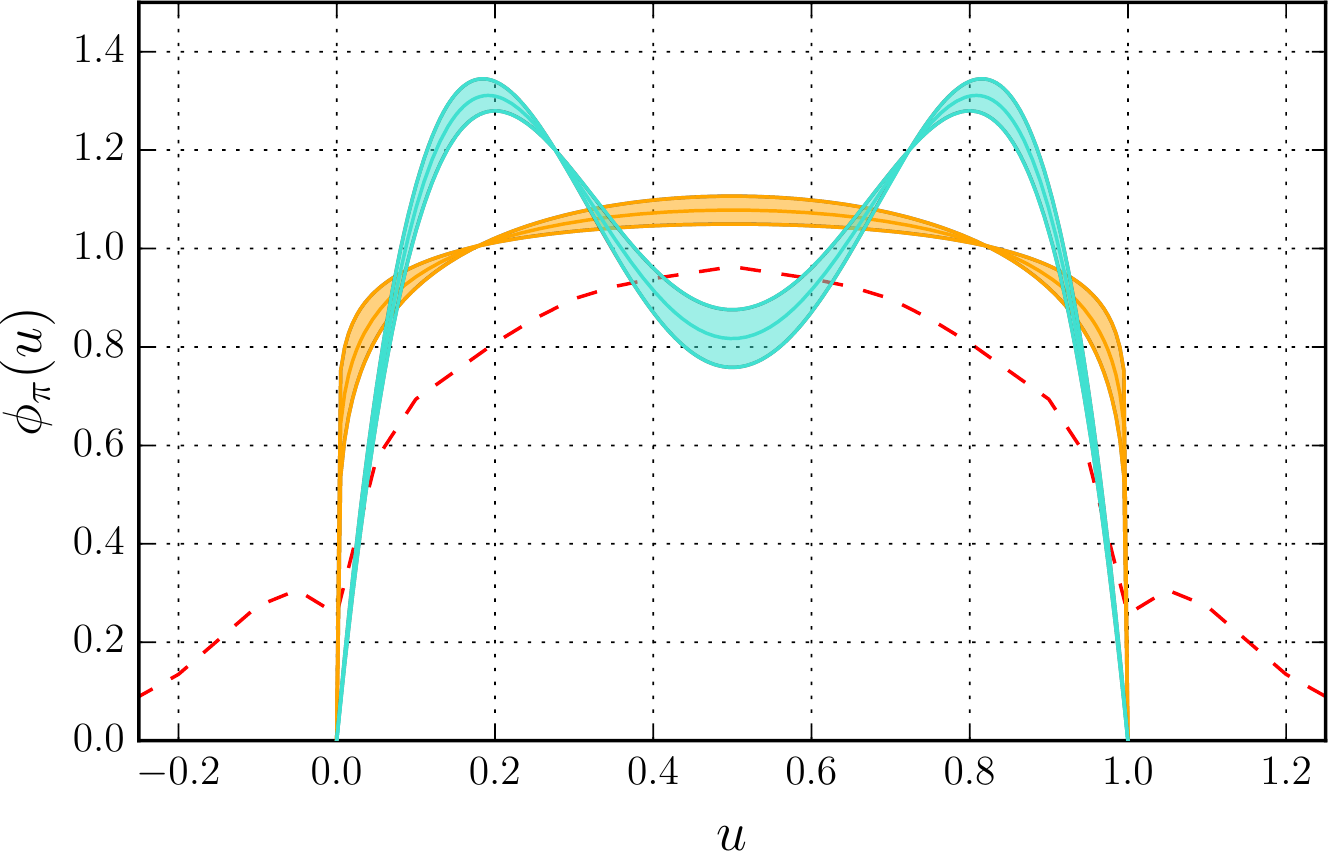}
\includegraphics[width=0.49\textwidth,keepaspectratio]{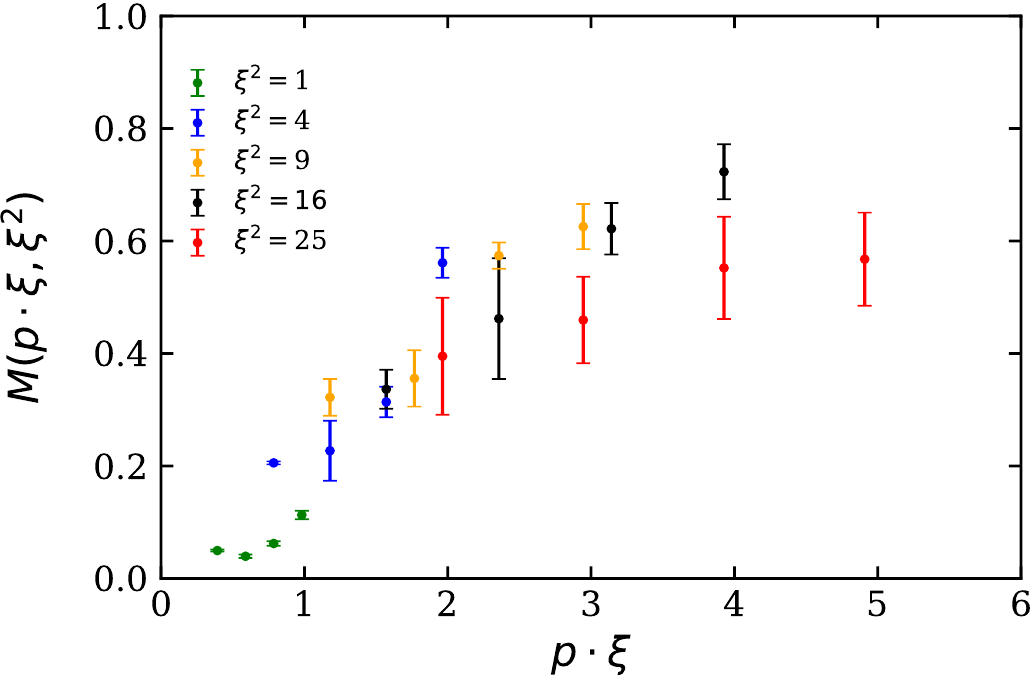}
\end{figure}

The first results for pion DAs were presented in \cite{Bali:2017gfr}, with a dedicated study of systematic uncertainties--particularly higher twist effects--appearing in \cite{Bali:2018spj}. The lattice data in \cite{Bali:2018spj} was obtained with six different pion momenta up to approximately $P\simeq \SI{2}{GeV}$, with ten current separations in the range $3.3a \leq z \leq 5a$, using a single $n_f = 2$ clover fermion ensemble with lattice spacing $a \simeq \SI{0.071}{fm}$, volume $L^3/a^3\times T/a = 32^3\times 64$, and $m_\pi \simeq \SI{295}{MeV}$. Fig.~\ref{fig:piDA} shows two fits to this lattice data, using two different model parameterisations and a comparison to the pion DA results from the quasi PDF approach \cite{Zhang:2017bzy}, indicated by the dashed red line.
The data are restricted to a small range of Ioffe times, for which both parameterisations lead to equally good descriptions of the results. More data are required to determine the DA and pin down higher twist effects.

The right hand plot in Fig.~\ref{fig:piDA} illustrates preliminary results, from \cite{Sufian:2018jla}, for the factorisable matrix element of two vector currents, as a function of the Ioffe time. Here $\zeta^2$ is the current separation in lattice units, and higher twist effects manifest in the splitting between different values of the current separation at fixed Ioffe time, visible particularly at large values of the current separation.

\section{\label{sec:hadront}Structure functions from lattice QCD}

So far I have focussed on the calculation of PDFs, but the direct computation of structure functions via the hadronic tensor or Compton amplitude offer complementary approaches. The differential cross-section for unpolarised DIS can be decomposed into leptonic and hadronic contributions, where the spin-averaged hadronic tensor is given by
\begin{align}
W_{\mu\nu} {} &(q,P) = \frac{1}{4\pi}\int \mathrm{d}^4y\,e^{iq\cdot y}\left\langle P | [j^\mu(y),j^\nu(0)] | P\right\rangle \nonumber \\
= {} & \left(-g^{\mu\nu} + \frac{q^\mu q^\nu}{q^2}\right)F_1(x,Q^2)
+ \frac{1}{P\cdot q}\left(P^\mu - q^\mu\frac{P\cdot q}{q^2}\right)\left(P^\nu - q^\nu\frac{P\cdot q}{q^2}\right)F_2(x,Q^2).
\end{align}
Here, in the second line, I have expressed the hadronic tensor in terms of the form factors of Sec.~\ref{sec:pdfs}. The optical theorem relates the hadronic tensor to the forward Compton amplitude\footnote{In particular, the discontinuity of the forward Compton amplitude across the cut starting at $\omega = 1$ in the complex $\omega$ plane gives the hadronic tensor for nucleon targets.}
\begin{align}
T_{\mu\nu}(q,P) = {} & i\int \mathrm{d}^4y\,e^{iq\cdot y}\left\langle P | {\cal T}\{j^\mu(y)j^\nu(0)\} | P\right\rangle \nonumber \\
= {} & \left(-g^{\mu\nu} + \frac{q^\mu q^\nu}{q^2}\right)\widetilde{F}_1(x,Q^2)
+ \frac{1}{P\cdot q}\left(P^\mu - q^\mu\frac{P\cdot q}{q^2}\right)\left(P^\nu - q^\nu\frac{P\cdot q}{q^2}\right)\widetilde{F}_2(x,Q^2),
\end{align}
where ${\cal T}$ is the time-ordering operator, and the corresponding structure functions are related by $\mathrm{Im}\,\widetilde{F}_{1,2}(\omega+i\epsilon) = 2\pi F_{1,2}(\omega)$, with $\omega = 1/x$.

\subsection{\label{ssec:liu}The Euclidean hadronic tensor}

Proposals to calculate the hadronic tensor in lattice QCD date back several decades \cite{Liu:1993cv}, but, until recently, have proved numerically too challenging to be computationally practical \cite{Liang:2017mye}. The starting point for this approach is the infinite volume Euclidean path integral formalism, from which one can extract the hadronic tensor in the Euclidean time representation
\begin{equation}
\widetilde{W}_{\mu\nu}(\mathbf{p},\mathbf{q},t_2-t_1) = \int \frac{\mathrm{d}^3\mathbf{y}}{2\pi}e^{-i\mathbf{q}\cdot\mathbf{y}}\left\langle P | J_\mu(\mathbf{y},t_2)J_\nu(0,t_1) | P\right\rangle,
\end{equation}
 via a ratio of four- and two-point functions. The analytic continuation to Minkowski spacetime is carried out via an inverse Laplace transform 
\begin{equation}
W_{\mu\nu}(Q^2,E) = \frac{1}{2iM}\int_{c-i\infty}^{c+i\infty}\mathrm{d}t\,e^{Et}\widetilde{W}_{\mu\nu}(\mathbf{p},\mathbf{q},t),
\end{equation}
where $c > 0$ and I denote the energy transfer by $E$ to distinguish it from the Ioffe time, $\nu$. 

Through this approach, a decomposition into valence and connected and disconnected sea quark topologies has been proposed \cite{Liu:1993cv}. The connected sea contribution is precisely that probed at negative $x$ in the quasi PDF approach, which can be interpreted through crossing symmetry as the antiquark contribution at positive $x$, and is responsible for the Gottfried sum rule violation.

\begin{figure}
\centering
\caption{\label{fig:hadE}(Left) The hadronic tensor from \cite{Liang:2017mye}. (Right) The Compton amplitude from \cite{Somfleth:2018msu}.}
\includegraphics[width=0.49\textwidth,keepaspectratio]{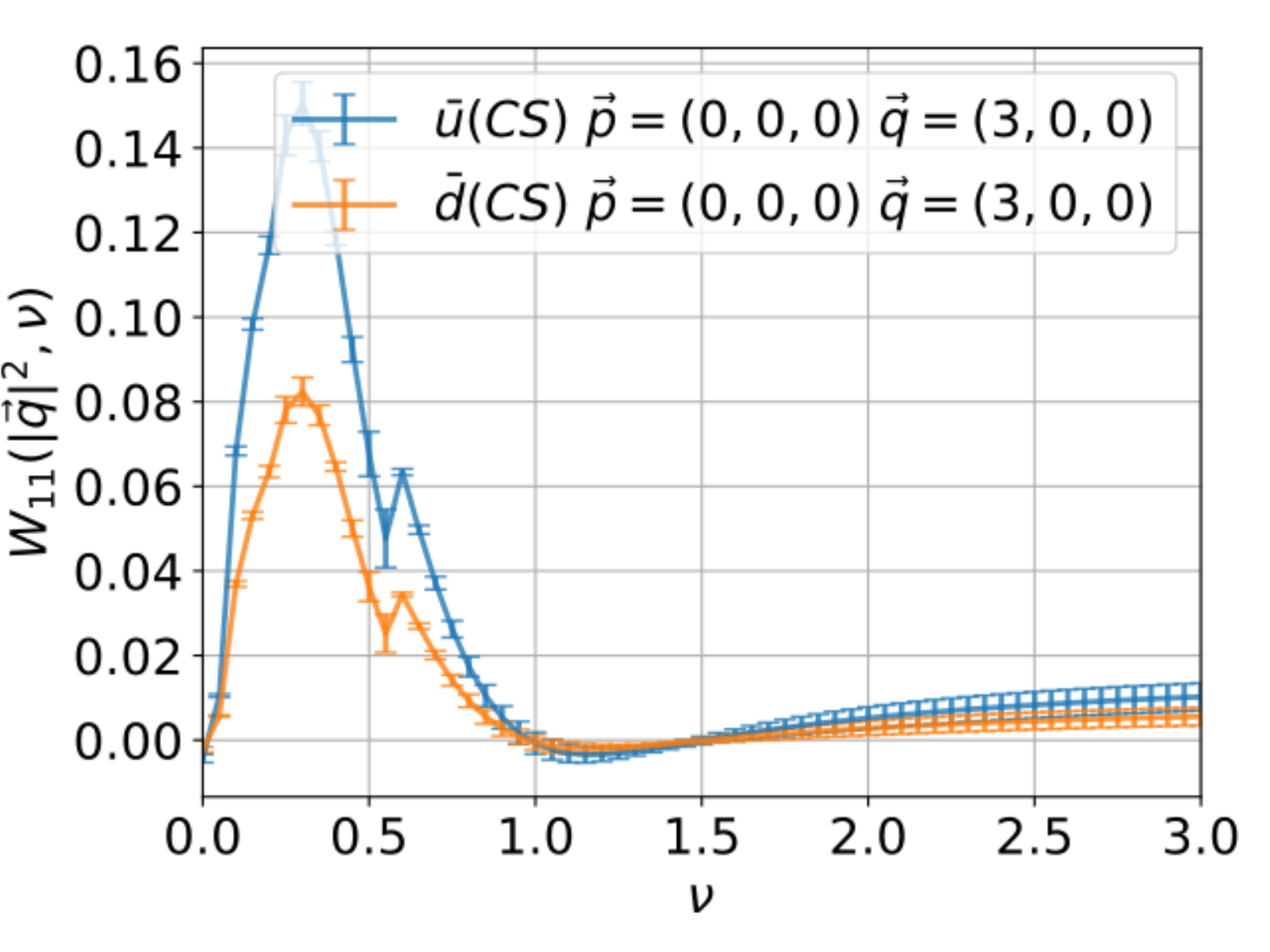}
\includegraphics[width=0.48\textwidth,keepaspectratio]{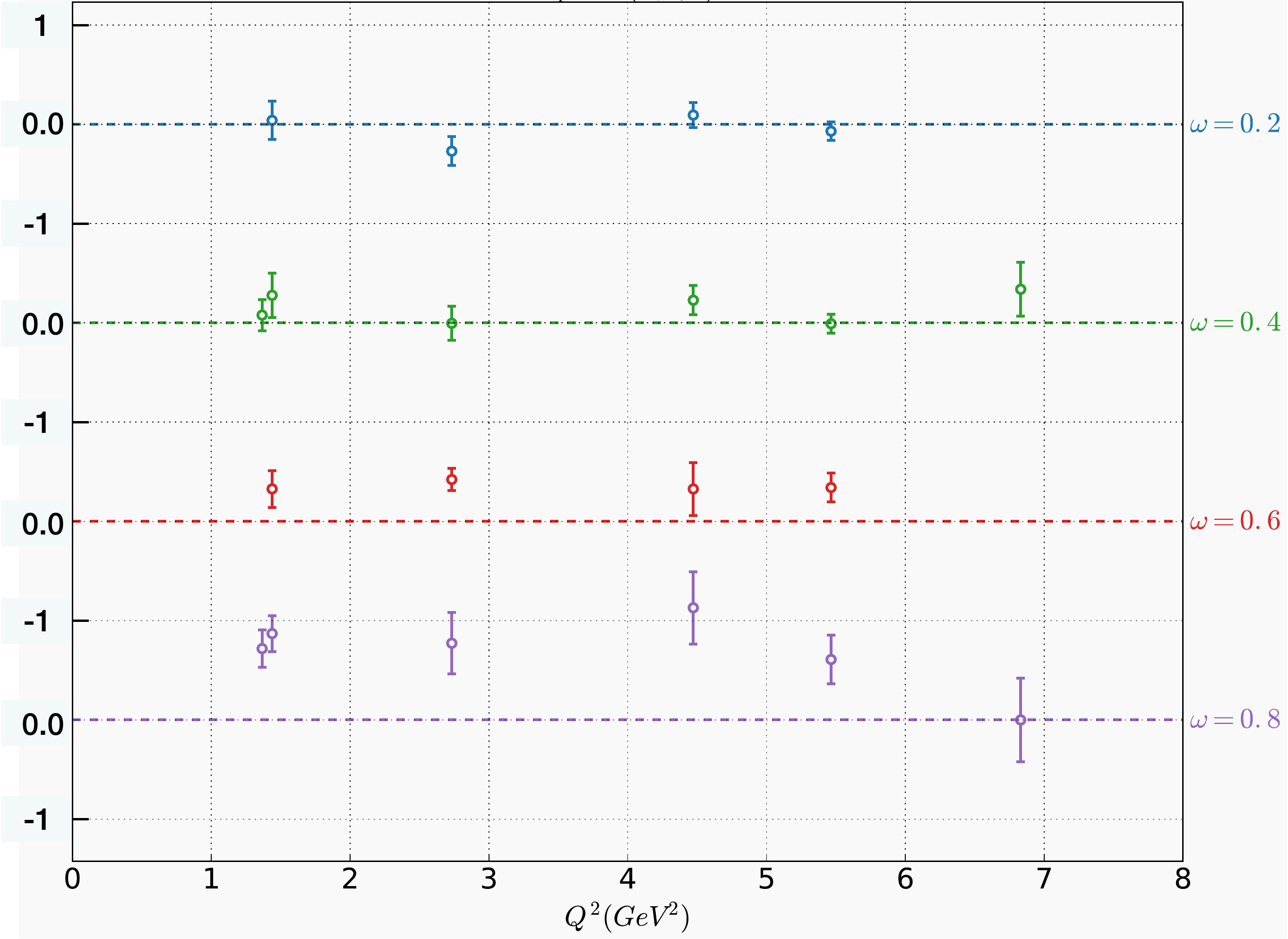}
\end{figure}

This approach is quite direct, but faces two challenges: the four-point function calculation is computationally expensive, and the inverse Laplace transform is a difficult inverse problem. The authors of \cite{Liang:2017mye} study the Backus-Gilbert reconstruction (used in \cite{Hansen:2017mnd}) with both toy model data and preliminary lattice data on a $L^3/a^3\times T/a = 12^3\times 128$ anisotropic lattice with spatial lattice spacing $a_s \simeq \SI{0.18}{fm}$ and $m_\pi = \SI{640}{MeV}$. The elastic and quasi elastic peaks are visible in the data, shown in the left hand plot in Fig.~\ref{fig:hadE}, but the DIS region is inaccessible. The blue lines show the reconstructed anti-up (``connected sea'') quark contribution and the orange line the corresponding results for the anti-down quark, for a specific choice of spatial momentum transfer $|\mathbf{q}|~\SI{1.7}{GeV}$ to a nucleon at rest, indicated by $|\mathbf{p}| = 0$. Larger momenta are required so that the energy transfer $E$ is smaller than the spatial momentum transfer, $|\mathbf{q}|$, and still sufficiently large that it is above the inelastic peak. It is also important to note that this discussion has been framed in infinite volume, and finite volume effects are intimately tied up with the Euclidean signature \cite{Hansen:2017mnd}.

\subsection{\label{ssec:qcdsf}The lattice OPE} 

The direct calculation of the forward Compton amplitude, via the Feynman-Hellmann method, was proposed in \cite{Chambers:2017dov}. First, by applying the OPE to the forward Compton amplitude, one can relate the structure functions $\widetilde{F}_{1,2}(\omega)$ to the Mellin moments of $F_{1,2}(\omega)$ via
\begin{align}
T_{\mu\nu}(q,P) = {} & \sum_{n=1}^\infty \bigg\{\left(-g^{\mu\nu} + \frac{q^\mu q^\nu}{q^2}\right)4\omega^{2n} \int_0^1\mathrm{d}x\,x^{2n-1}F_1(x,Q^2) \nonumber\\
{} & + \frac{1}{P\cdot q}\left(P^\mu - q^\mu\frac{P\cdot q}{q^2}\right)\left(P^\nu - q^\nu\frac{P\cdot q}{q^2}\right)4\omega^{2n-1} \int_0^1\mathrm{d}x\,x^{2n-2}F_2(x,Q^2)\bigg\}.
\end{align}
Then, summing the geometric series, one can identify specific components of the forward Compton amplitude with individual structure functions. For example, taking $\mu = \nu = 3$ and $P_3 = q_3 = q_4 = 0$, we have
\begin{equation}
T_{33} = 4\omega \int_0^1\mathrm{d}x\frac{\omega x}{1-\omega^2x^2}F_1(x,Q^2),
\end{equation}
which can also be derived using a dispersive approach. 

The relevant component of the forward Compton amplitude can be determined on the lattice rather efficiently via an extension of the Feynman-Hellman method, which relates matrix elements to variations in the spectrum. In particular, by taking the second derivative of the ground state energy in the presence of a source term $J_\mu(x) = \lambda Z_V \cos(\mathbf{q}\cdot \mathbf{x}) e_f \overline{\psi}(x)\gamma_\mu \psi_f(x)$, where $e_f$ is the electric charge of a quark of flavour $f$, the Compton tensor can be extracted from $T_{33} = -2E_\lambda \partial^2E_\lambda /\partial\lambda^2 |_{\lambda = 0}$,
where $E_\lambda$ is the ground state energy of the nucleon in the presence of the source term $J_\mu(x)$. This method avoids the need to compute a four point function directly, at the expense of repeated calculations at different values of $\lambda$.

Preliminary results for the Compton amplitude, taken from \cite{Somfleth:2018msu}, are shown in the right panel of Fig.~\ref{fig:hadE}. The amplitude is plotted as a function of the momentum transfer $Q^2$, at four different values of $\omega$; the amplitude is largely independent of $Q^2$ at fixed $\omega$, demonstrating the lattice analogue of Bjorken scaling of the structure functions.

\section{\label{sec:many}``Many moment methods''}

I have so far focussed on methods aimed at the direct calculation of the $x$ dependence of PDFs, DAs, and structure functions. I close by briefly highlighting two approaches\footnote{The direct determination of the Compton amplitude can also be construed as a ``many moments method'', if the geometric summation is not carried out.} that I dub ``many moments methods''.

One long-standing approach to lattice hadron structure calculations is to calculate the lowest Mellin moments of PDFs (which include, for example, the tensor charge of the nucleon \cite{Alexandrou:2017qyt}). Mellin moments of PDFs can be related to matrix elements of local twist-two operators, where twist is the dimension minus the spin of the operator, that can be calculated using traditional lattice QCD techniques. In principle, given a sufficient number of Mellin moments of a function, the inverse Mellin transform can be approximated numerically and the original function can be reconstructed. In practice, however, the hypercubic symmetry of the lattice regulator induces power-divergent mixing between operators of different mass dimension, because rotational symmetry is broken and angular momentum is no longer a good quantum number. Power-divergent mixing limits lattice calculations to the lowest three Mellin moments of PDFs\footnote{The lowest moments of PDFs do provide useful information for phenomenological determinations of PDFs \cite{Lin:2017snn}, however, particularly for the cases for which experimental data are scarce. For example, the tensor charge significantly constrains the shape of the transversity distribution, as shown in Fig.~\ref{fig:qpdf_etmc} (see also \cite{Lin:2017stx}).}, which is insufficient to reconstruct PDFs without significant model input \cite{Detmold:2001dv}. 

In light of the success of the $x$-dependent methods sketched above, two approaches that circumvent power divergent mixing are now being revisited. The first  proposes using flavour-changing currents that couple light quarks to a fictitious heavy quark to construct the (spin-averaged) heavy-light Compton tensor \cite{Detmold:2005gg}
\begin{equation}
T_{[\mu\nu]}^{(\Psi,\psi)}(q,P) = \int\mathrm{d}^4y\, e^{iq\cdot y} \langle 0| {\cal T}\{A_\mu^{(\Psi,\psi)}(y)A_\nu^{(\Psi,\psi)}(0)|\pi(P)\rangle,
\end{equation}
with Lorentz indices antisymmetrised and $A_\mu^{(\Psi,\psi)}(y) = \overline{\Psi}(y)\gamma_\mu \gamma_5\psi(y) + \overline{\psi}(y)\gamma_\mu \gamma_5 \Psi(y)$. Then one can show that \cite{Detmold:2005gg}
\begin{equation}
T_{[\mu\nu]}^{(\Psi,\psi)} = 2if_\pi \epsilon_{\mu\nu\rho\sigma}q^\rho P^\sigma\sum_{n=0}^\infty a_{2n}^{\psi}(\mu^2)f_n.
\end{equation}
Here the $a_{2n}^{\psi}$ are the Mellin moments of the pion DA, and the $f_n$ functions are known functions of kinematic variables and Wilson coefficients. This approach requires the the continuum limit to be taken, and that the Minkowski momenta are in the unphysical region $(q_M + P_M)^2 < (m_\Psi + \Lambda_{\mathrm{QCD}})^2$, where $q_M$ and $P_M$ are the Minkowski counterparts of $q$ and $P$. Higher twist contributions can be extracted from the off-diagonal elements of the heavy-light Compton tensor, and PDFs can be determined via heavy-light vector currents. This method requires very fine lattices in the temporal direction to avoid large heavy quark discretisation effects, such that
$\Lambda_{\mathrm{QCD}} \ll \sqrt{q^2} \lesssim m_\Psi \ll 1/a$. 

The first preliminary quenched results for the heavy-light hadronic tensor are shown in the right hand plot in Fig.~\ref{fig:ppdf}, from \cite{Detmold:2018kwu}, as a function of the lattice spacing, $a$, for four different values of the momentum $q$, with $\mathbf{q}=(0,0,2\pi/L)$ and $\mathbf{P} = \mathbf{0}$ fixed. The lattice spacings correspond to $a=\SI{0.05}{fm}$, $\SI{0.06}{fm}$, and $\SI{0.075}{fm}$, with box sizes of $L^3/a^3\times T/a=48^3\times96$, $40^3\times 80$, and $32^3\times64$, respectively. The valence quarks are nonperturbatively improved Wilson clover quarks, with heavy quark mass $m_\Psi = \SI{1.3}{GeV}$, and the currents are renormalised via one loop perturbation theory. Discretisation effects are clearly visible in these results, and more data are required to extract information on the pion DA, which requires the continuum limit. 

An alternative many moments method, using smeared operators of fixed physical size, was proposed in \cite{Davoudi:2012ya} and is currently being studied nonperturbatively. By fixing the size of the operators in physical units, the power-divergent mixing is reduced to lattice artefacts that vanish in the continuum limit. One particularly natural way to implement the smearing is the gradient flow \cite{Monahan:2015lha}.

\section{\label{sec:outro}Outlook}

The direct determination of $x$-dependent hadron structure from first principles has been a long-standing challenge for lattice QCD. Although there are still obstacles to overcome, and some of the theoretical landscape is yet to be explored, the last five years have seen significant progress in sketching out a map for the route forward. A signature milestone was finally achieved this year: the first lattice calculations of PDFs at physical pion masses.

Quasi and pseudo PDFs have been studied extensively, and the relationship between the underlying Ioffe time matrix element and the three distribution functions has been clarified through factorisation theorems. Nonperturbative results, involving a completely nonperturbative renormalisation procedure, for both mesons and nucleons have now appeared, although generally only at a single lattice spacing, volume, and pion mass. In fact, significant work is required to understand and quantify the complete set of systematic uncertainties associated with the determination of PDFs from quasi and pseudo PDF calculations. 

Inspired, at least in part, by the obvious success of the quasi and pseudo PDF programs, there have been new studies of a number of older suggestions, including the direct calculation of the hadronic tensors and methods to calculate higher moments of PDFs and DAs, and new ideas for the direct determination of the Compton amplitude. Moreover, the computation of distribution functions in real space has been revisited, and a generalised, overarching approach, akin to the global extraction of PDFs from experimental data, has been proposed, which goes by the catchy misnomer ``good lattice cross-sections'', but is more correctly understood as ``factorisable matrix elements''. Preliminary nonperturbative results have started to appear over the last year, but much work remains. 

These approaches each have theoretical advantages and individual challenges, but all offer complementary paths to the same underlying hadron structure. We may not have entered the era of precision nonperturbative calculations of $x$-dependent hadron structure, but it is clear that the new era dawns.

\begin{acknowledgments}
I thank the following for willingly sharing their results and discussing their work: Bipasha Chakraborty, Krzysztof Cichy, Martha Constantinou, Michael Engelhardt, Nikhil Karthik, Joseph Karpie, Jian Liang, David Lin, Keh-Feh Liu, Yu-Sheng Liu, Santanu Mondal, Giancarlo Rossi, Aurora Scapellato, Kim Somfleth, Raza Sufian, Massimo Testa, Philipp Wein, Yi-Bo Yang, Savvas Zafeiropoulos, James Zanotti, Jianhui Zhang, and Yong Zhao. I am particularly grateful to Kostas Orginos for our many discussions of hadron structure. I am supported in part
by the U.S.~Department of Energy through Grant No.~DE-FG02-00ER41132.
\end{acknowledgments}
\vspace*{-0.5\baselineskip}


\begin{thebibliography}{99}
\setlength{\itemsep}{-2pt}

\vspace*{-0.5\baselineskip}
\begin{footnotesize}
\bibitem{Gao:2017yyd}
J.~Gao \emph{et al.}, \emph{Phys.~Rept.~}{\bf 742} (2018) 1
[{\tt \href{http://arxiv.org/abs/1709.04922}{1709.04922}}]; 
L.~Del Debbio, \emph{EPJ Web Conf.}~{\bf 175} (2018) 01006

\bibitem{Collins:2011zzd}
J.~Collins, \emph{Foundations of Perturbative QCD}, Cambridge University Press, 2011

\bibitem{Lin:2017snn}
H.-W.~Lin \emph{et al.}, \emph{Prog.~Part.~Nucl.~Phys.~}{\bf 100} (2018) 107
[{\tt \href{http://arxiv.org/abs/1711.07916}{1711.07916}}]

\bibitem{Liu:2018uuj}
Y.-S.~Liu \emph{et al.}, {\tt \href{http://arxiv.org/abs/1807.06566}{1807.06566}}

\bibitem{Cichy:2018int}
K.~Cichy \emph{et al.}, \pos{PoS(LATTICE2018)094}; C.~Alexandrou \emph{et al.}, \emph{In preparation}

\bibitem{Braun:2018brg}
V.~Braun \emph{et al.}, {\tt \href{http://arxiv.org/abs/1810.00048}{1810.00048}}

\bibitem{Bali:2018spj}
G.~Bali \emph{et al.}, {\tt \href{http://arxiv.org/abs/1807.06671}{1807.06671}}

\bibitem{Moffat:2017sha}
E.~Moffat \emph{et al.}, \emph{Phys.~Rev.~D}{\bf 95} (2017) 096008
[{\tt \href{http://arxiv.org/abs/1702.03955}{1702.03955}}]

\bibitem{Braun:1994jq}
B.~Ioffe, \emph{Phys.~Lett.~B} {\bf 30} (1969) 123;
V.~Braun \emph{et al.}, \emph{Phys.~Rev.~D} {\bf 51} (1995) 6036
[{\tt \href{http://arxiv.org/abs/hep-ph/9410318}{hep-ph/9410318}}]

\bibitem{Ji:2013dva}
X.~Ji, \emph{Phys.~Rev.~Lett.~}{\bf 26} (2013) 262002 
[{\tt \href{http://arxiv.org/abs/1305.1539}{1305.1539}}]

\bibitem{Radyushkin:2017cyf}
A.~Radyushkin, \emph{Phys.~Rev.~D} {\bf 96} (2017) 034025
[{\tt \href{http://arxiv.org/abs/1705.01488}{1705.01488}}]

\bibitem{Musch:2010ka}
B.~Musch \emph{et al.}, \emph{Phys.~Rev.~D} {\bf 83} (2011) 094507
[{\tt \href{http://arxiv.org/abs/1011.1213}{1011.1213}}]; B.~Yoon \emph{et al.}, \emph{Phys.~Rev.~D} {\bf 96} (2017) 094508
[{\tt \href{http://arxiv.org/abs/1706.03406}{1706.03406}}]

\bibitem{Briceno:2017cpo}
R.~Brice{\~n}o \emph{et al.}, \emph{Phys.~Rev.~D} {\bf 96} (2017) 014502
[{\tt \href{http://arxiv.org/abs/1703.06072}{1703.06072}}]

\bibitem{Ji:2017rah}
X.~Ji \emph{et al.}, \emph{Nucl.~Phys.~B} {\bf 924} (2017) 366 
[{\tt \href{http://arxiv.org/abs/1706.07416}{1706.07416}}]

\bibitem{Izubuchi:2018srq}
T.~Izubuchi \emph{et al.}, \emph{Phys.~Rev.~D} {\bf 98} (2018) 056004
[{\tt \href{http://arxiv.org/abs/1801.03917}{1801.03917}}];
J.-H.~Zhang \emph{et al.}, \emph{Phys.~Rev.~D} {\bf 97} (2018) 074508
[{\tt \href{http://arxiv.org/abs/1801.03023}{1801.03023}}];
A.~Radyushkin, \emph{Phys.~Rev.~D} {\bf 98} (2018) 014019
[{\tt \href{http://arxiv.org/abs/1801.02427}{1801.02427}}]

\bibitem{Radyushkin:2018nbf}
A.~Radyushkin, {\tt \href{http://arxiv.org/abs/1807.07509}{1807.07509}}

\bibitem{Radyushkin:2016hsy}
A.~Radyushkin, \emph{Phys.~Lett.~B} {\bf 767} (2017) 314
[{\tt \href{http://arxiv.org/abs/1612.05170}{1612.05170}}]

\bibitem{Ji:2014gla}
X.~Ji, \emph{Sci.~China Phys.~Mech.~Astron.~}{\bf 57} (2014) 1407
[{\tt \href{http://arxiv.org/abs/1404.6680}{1404.6680}}]

\bibitem{Parisi:1984ggi}
G.~Parisi, \emph{Phys.~Rept.~}{\bf 103} (1984) 2013; 
G.~P.~Lepage, \emph{Boulder ASI Lectures} (1989) 97

\bibitem{Bali:2016lva}
G.~Bali \emph{et al.}, \emph{Phys.~Rev.~D} {\bf 93} (2016) 094515
[{\tt \href{http://arxiv.org/abs/1602.05525}{1602.05525}}]

\bibitem{Gamberg:2014zwa}
L.~Gamberg \emph{et al.}, \emph{Phys.~Lett.~B} {\bf 743} (2015) 112
[{\tt \href{http://arxiv.org/abs/1412.3401}{1412.3401}}];
W.~Broniowski and E.~Ruiz Arriola, \emph{Phys.~Lett.~B} {\bf 773} (2017) 385
[{\tt \href{http://arxiv.org/abs/1707.09588}{1707.09588}}];
T.~Hobbs, \emph{Phys.~Rev.~D} {\bf 97} (2018) 054028
[{\tt \href{http://arxiv.org/abs/1708.05463}{1708.05463}}]

\bibitem{Ishikawa:2017faj}
T.~Ishikawa \emph{et al.},  \emph{Phys.~Rev.~D} {\bf 96} (2017) 094019
[{\tt \href{http://arxiv.org/abs/1707.03107}{1707.03107}}];
X.~Ji \emph{et al.}, \emph{Phys.~Rev.~Lett.~}{\bf 120} (2018) 112001
[{\tt \href{http://arxiv.org/abs/1706.08962}{1706.08962}}];
J.~Green \emph{et al.}, \emph{Phys.~Rev.~Lett.~}{\bf 121} (2018) 022004
[{\tt \href{http://arxiv.org/abs/1707.07152}{1707.07152}}]

\bibitem{Zhang:2018diq}
J.-H.~Zhang \emph{et al.}, {\tt \href{http://arxiv.org/abs/1808.10824}{1808.10824}};
Z.-Y.~Li \emph{et al.}, {\tt \href{http://arxiv.org/abs/1809.01836}{1809.01836}}

\bibitem{Constantinou:2017sej}
M.~Constantinou, and H.~Panagopoulos, \emph{Phys.~Rev.~D} {\bf 96} (2017) 054506
[{\tt \href{http://arxiv.org/abs/1705.11193}{1705.11193}}]

\bibitem{Chen:2017mie}
J.-W.~Chen \emph{et al.}, {\tt \href{http://arxiv.org/abs/1710.01089}{1710.01089}}

\bibitem{Rossi:2018zkn}
G.~Rossi and M.~Testa, \emph{Phys.~Rev.~D} {\bf 98} (2018) 054028
[{\tt \href{http://arxiv.org/abs/1806.00808}{1806.00808}}];
G.~Rossi and M.~Testa, \emph{Phys.~Rev.~D} {\bf 96} (2017) 014507
[{\tt \href{http://arxiv.org/abs/1706.04428}{1706.04428}}]

\bibitem{Karpie:2018zaz}
J.~Karpie \emph{et al.}, {\tt \href{http://arxiv.org/abs/1807.10933}{1807.10933}}

\bibitem{Xiong:2017jtn}
X.~Xiong \emph{et al.}, {\tt \href{http://arxiv.org/abs/1705.00246}{1705.00246}}

\bibitem{Ji:2015jwa}
X.~Ji and J.-H.~Zhang, \emph{Phys.~Rev.~D} {\bf 92} (2015) 034006
[{\tt \href{http://arxiv.org/abs/1505.07699}{1505.07699}}]

\bibitem{Monahan:2016bvm}
C.~Monahan and K.~Orginos, \emph{JHEP} {\bf 1603} (2017) 116
[{\tt \href{http://arxiv.org/abs/1612.01584}{1612.01584}}];
C.~Monahan, \emph{Phys.~Rev.~D} {\bf 97} (2018) 054507
[{\tt \href{http://arxiv.org/abs/1710.04607}{1710.04607}}]

\bibitem{Alexandrou:2017huk}
C.~Alexandrou \emph{et al.}, \emph{Nucl.~Phys.~B} {\bf 923} (2017) 394
[{\tt \href{http://arxiv.org/abs/1706.00265}{1706.00265}}];
J.-W.~Chen \emph{et al.}, \emph{Phys.~Rev.~D} {\bf 97} (2018) 014505
[{\tt \href{http://arxiv.org/abs/1706.01295}{1706.01295}}]

\bibitem{Spanoudes:2018zya}
G.~Spanoudes and H.~Panagopoulos, \emph{Phys.~Rev.~D} {\bf 98} (2018) 014509
[{\tt \href{http://arxiv.org/abs/1805.01164}{1805.01164}}]

\bibitem{Liu:2018tox}
Y.-S.~Liu \emph{et al.}, {\tt \href{http://arxiv.org/abs/1810.10879}{1810.10879}}

\bibitem{Stewart:2017tvs}
I.~Stewart and Y.~Zhao, \emph{Phys.~Rev.~D} {\bf 97} (2018) 054512
[{\tt \href{http://arxiv.org/abs/1709.04933}{1709.04933}}]

\bibitem{Alexandrou:2018pbm}
C.~Alexandrou \emph{et al.}, \emph{Phys.~Rev.~Lett.~}{\bf 121} (2018) 112001
[{\tt \href{http://arxiv.org/abs/1803.02685}{1803.02685}}]

\bibitem{Xiong:2013bka}
X.~Xiong \emph{et al.}, \emph{Phys.~Rev.~D} {\bf 90} (2014) 014051
[{\tt \href{http://arxiv.org/abs/1310.7471}{1310.7471}}]

\bibitem{Ma:2014jla}
Y.-Q.~Ma and J.-W~Qiu, \emph{Int.~J.~Mod.~Phys.~Conf.~Ser.~}{\bf 37} (2015) 1560041
[{\tt \href{http://arxiv.org/abs/1412.2688}{1412.2688}}];
Y.-Q.~Ma and J.-W~Qiu, \emph{Phys.~Rev.~D} {\bf 98} (2018) 074021
[{\tt \href{http://arxiv.org/abs/1404.6860}{1404.6860}}];
Y.-Q.~Ma and J.-W~Qiu, \emph{Phys.~Rev.~Lett.~}{\bf 120} (2018) 022003
[{\tt \href{http://arxiv.org/abs/1709.03018}{1709.03018}}]

\bibitem{Dotsenko:1979wb}
V.~Dotsenko and S.~Vergeles, \emph{Nucl.~Phys.~B} {\bf 169} (1980) 527;
N.~Craigie and H.~Dorn, \emph{Nucl.~Phys.~B} {\bf 185} (1981) 204;
H.~Dorn, \emph{Fortsch.~Phys.~}{\bf 34} (1986) 11

\bibitem{Ishikawa:2016znu}
T.~Ishikawa \emph{et al.}, {\tt \href{http://arxiv.org/abs/1609.02018}{1609.02018}};
J.-W.~Chen \emph{et al.}, \emph{Nucl.~Phys.~B} {\bf 915} (2017) 1
[{\tt \href{http://arxiv.org/abs/1609.08102}{1609.08102}}]

\bibitem{Martinelli:1994ty}
G.~Martinelli \emph{et al.}, \emph{Nucl.~Phys.~B} {\bf 445} (1995) 81
[{\tt \href{http://arxiv.org/abs/hep-lat/9411010}{hep-lat/9411010}}]

\bibitem{Orginos:2017kos}
K.~Orginos \emph{et al.}, \emph{Phys.~Rev.~D} {\bf 96} (2017) 094503
[{\tt \href{http://arxiv.org/abs/1706.05373}{1706.05373}}];
J.~Karpie \emph{et al.}, \emph{In preparation}

\bibitem{Narayanan:2006rf}
R.~Narayanan, and H.~Neuberger, \emph{JHEP} {\bf 0603} (2006) 064
[{\tt \href{http://arxiv.org/abs/hep-th/0601210}{hep-th/0601210}}];
M.~{L\"uscher}, \emph{JHEP} {\bf 1008} (2010) 071
[{\tt \href{http://arxiv.org/abs/1006.4518}{1006.4518}}]

\bibitem{Luscher:2011bx}
M.~{L\"uscher, M.}, \emph{JHEP} {\bf 1304} (2013) 123
[{\tt \href{http://arxiv.org/abs/1302.5246}{1302.5246}}];
M.~{L\"uscher, M.} and P.~Weisz, \emph{JHEP} {\bf 1102} (2011) 051
[{\tt \href{http://arxiv.org/abs/1101.0963}{1101.0963}}]; 
M.~{L\"uscher}, \emph{JHEP} {\bf 1008} (2010) 071
[{\tt \href{http://arxiv.org/abs/1006.4518}{1006.4518}}]

\bibitem{Dudek:2009qf}
J.~Dudek \emph{et al.}, \emph{Phys.~Rev.~Lett.~}{\bf 103} (2009) 262001
[{\tt \href{http://arxiv.org/abs/0909.0200}{0909.0200}}];
J.~Dudek \emph{et al.}, \emph{Phys.~Rev.~D}{\bf 82} (2010) 034508
[{\tt \href{http://arxiv.org/abs/1004.4930}{1004.4930}}]

\bibitem{Briceno:2018lfj}
R.~Brice{\~n}o \emph{et al.}, \emph{Phys.~Rev.~D} {\bf 98} (2018) 014511
[{\tt \href{http://arxiv.org/abs/1805.01034}{1805.01034}}];
R.~Brice{\~n}o \emph{et al.}, \pos{PoS(LATTICE2018)111} 

\bibitem{Lin:2017ani}
H.-W.~Lin \emph{et al.}, \emph{Phys.~Rev.~D} {\bf 98} (2018) 054504
[{\tt \href{http://arxiv.org/abs/1708.05301}{1708.05301}}];
J.-W.~Chen \emph{et al.}, (2017) {\tt \href{http://arxiv.org/abs/1711.07858}{1711.07858}};

\bibitem{Zafeiropoulos:2018lat}
J.~Karpie \emph{et al.}, \pos{PoS(LATTICE2018)100}

\bibitem{Chen:2016utp}
J.-W.~Chen \emph{et al.}, \emph{Nucl.~Phys.~B} {\bf 911} (2016) 246
[{\tt \href{http://arxiv.org/abs/1603.06664}{1603.06664}}]

\bibitem{Lin:2014zya}
H.-W.~Lin \emph{et al.}, \emph{Phys.~Rev.~D} {\bf 91} (2015) 054510
[{\tt \href{http://arxiv.org/abs/1402.1462}{1402.1462}}];
C.~Alexandrou \emph{et al.}, \emph{Phys.~Rev.~D} {\bf 92} (2015) 014502
[{\tt \href{http://arxiv.org/abs/1504.07455}{1504.07455}}]

\bibitem{Alexandrou:2016jqi}
C.~Alexandrou \emph{et al.}, \emph{Phys.~Rev.~D} {\bf 96} (2017) 014513
[{\tt \href{http://arxiv.org/abs/1610.03689}{1610.03689}}];
C.~Alexandrou \emph{et al.}, \emph{Eur.~Phys.~J.~Web Conf.~}{\bf 175} (2018) 06021
[{\tt \href{http://arxiv.org/abs/1709.07513}{1709.07513}}]

\bibitem{Chen:2018xof}
J.-W.~Chen \emph{et al.},{\tt \href{http://arxiv.org/abs/1803.04393}{1803.04393}}

\bibitem{Lin:2018qky}
H.-W.~Lin \emph{et al.}, {\tt \href{http://arxiv.org/abs/1807.07431}{1807.07431}}

\bibitem{Alexandrou:2018eet}
C.~Alexandrou \emph{et al.}, accepted in \emph{Phys.~Rev.~D (Rapid Commun.)} [{\tt \href{http://arxiv.org/abs/1807.00232}{1807.00232}}]

\bibitem{Alexandrou:2017qyt}
C.~Alexandrou \emph{et al.}, \emph{Phys.~Rev.~D} {\bf 95} (2017) 114514
[{\tt \href{http://arxiv.org/abs/1703.08788}{1703.08788}}];
R.~Gupta \emph{et al.}, \emph{Phys.~Rev.~D} {\bf 98} (2018) 034503
[{\tt \href{http://arxiv.org/abs/1806.09006}{1806.09006}}]

\bibitem{Liu:2018hxv}
Y.-S.~Liu \emph{et al.}, {\tt \href{http://arxiv.org/abs/1810.05043}{1810.05043}}

\bibitem{Wang:2017qyg}
W.~Wang \emph{et al.}, \emph{Eur.~Phys.~J.~C} {\bf 78} (2018) 147
[{\tt \href{http://arxiv.org/abs/1708.02458}{1708.02458}}];
W.~Wang and S.~Zhao, \emph{JHEP} {\bf 05} (2018) 142
[{\tt \href{http://arxiv.org/abs/1712.09247}{1712.09247}}];
Z.-Y.~Fan \emph{et al.}, {\tt \href{http://arxiv.org/abs/1808.02077}{1808.02077}}

\bibitem{Zhang:2017bzy}
J.-H.~Zhang \emph{et al.}, \emph{Phys.~Rev.~D} {\bf 95} (2017) 094514
[{\tt \href{http://arxiv.org/abs/1702.00008}{1702.00008}}]

\bibitem{Chen:2018fwa}
J.-W.~Chen \emph{et al.}, {\tt \href{http://arxiv.org/abs/1804.01483}{1804.01483}};
N.~Karthik \emph{et al.}, \emph{In preparation}

\bibitem{Chen:2017gck}
J.-W.~Chen \emph{et al.}, {\tt \href{http://arxiv.org/abs/1712.10025}{1712.10025}}

\bibitem{Braun:2007wv}
V.~Braun and D.~Mueller, \emph{Eur.~Phys.~J.~C} {\bf 55} (2008) 349
[{\tt \href{http://arxiv.org/abs/hep-ph/0709.1348}{hep-ph/0709.1348}}]

\bibitem{Bali:2017gfr}
G.~Bali \emph{et al.}, \emph{Eur.~Phys.~J.~C} {\bf 78} (2018) 217
[{\tt \href{http://arxiv.org/abs/1709.04325}{1709.04325}}]

\bibitem{Sufian:2018jla}
R.~Sufian \emph{et al.}, \emph{In preparation}

\bibitem{Liu:1993cv}
K.-F.~Liu and S.-J.~Dong, \emph{Phys.~Rev.~Lett.~}{\bf 72} (1994) 1790
[{\tt \href{http://arxiv.org/abs/hep-ph/9306299}{hep-ph/9306299}}];
K.-F.~Liu, \emph{Phys.~Rev.~D}{\bf 62} (2000) 074501
[{\tt \href{http://arxiv.org/abs/hep-ph/9910306}{hep-ph/9910306}}];
K.-F.~Liu, \pos{PoS(LATTICE2015)115} 
[{\tt \href{http://arxiv.org/abs/1603.07352}{1603.07352}}];
K.-F.~Liu, \emph{Phys.~Rev.~D}{\bf 96} (2017) 033001
[{\tt \href{http://arxiv.org/abs/1703.04690}{1703.04690}}]

\bibitem{Liang:2017mye}
J.~Liang \emph{et al.}, \emph{Eur.~Phys.~J.~Web.~Conf.~}{\bf 175} (2018) 14014
[{\tt \href{http://arxiv.org/abs/1710.11145}{1710.11145}}]

\bibitem{Hansen:2017mnd}
M.~Hansen \emph{et al.}, \emph{Phys.~Rev.~D} {\bf 96} (2017) 094513
[{\tt \href{http://arxiv.org/abs/1704.08993}{1704.08993}}]

\bibitem{Chambers:2017dov}
A.~Chambers \emph{et al.}, \emph{Phys.~Rev.~Lett.~}{\bf 118} (2017) 242001
[{\tt \href{http://arxiv.org/abs/1703.01153}{1703.01153}}]

\bibitem{Somfleth:2018msu}
K.~Somfleth \emph{et al.}, \emph{In preparation}

\bibitem{Lin:2017stx}
H.-W.~Lin \emph{et al.}, \emph{Phys.~Rev.~Lett.~}{\bf 120} (2018) 152502 
[{\tt \href{http://arxiv.org/abs/1710.09858}{1710.09858}}]

\bibitem{Detmold:2001dv}
W.~Detmold \emph{et al.}, \emph{Eur.~Phys.~J.~direct C} {\bf 3} (2001) 1
[{\tt \href{http://arxiv.org/abs/hep-lat/0108002}{hep-lat/0108002}}];
W.~Detmold \emph{et al.}, \emph{Phys.Rev.~D} {\bf 68} (2003) 034025
[{\tt \href{http://arxiv.org/abs/hep-lat/0303015}{hep-lat/0303015}}];
W.~Detmold \emph{et al.}, \emph{Mod.~Phys.~Lett.~A} {\bf 18} (2003) 2681
[{\tt \href{http://arxiv.org/abs/hep-lat/0310003}{hep-lat/0310003}}]

\bibitem{Detmold:2005gg}
W.~Detmold and C.~Lin, \emph{Phys.~Rev.~D} {\bf 73} (2006) 014501
[{\tt \href{http://arxiv.org/abs/hep-lat/0507007}{hep-lat/0507007}}]

\bibitem{Detmold:2018kwu}
W.~Detmold \emph{et al.}, \pos{PoS(LATTICE2018)} 
[{\tt \href{http://arxiv.org/abs/1810.12194}{1810.12194}}]

\bibitem{Davoudi:2012ya}
Z.~Davoudi and M.~Savage, \emph{Phys.~Rev.~D} {\bf 86} (2012) 054505
[{\tt \href{http://arxiv.org/abs/1204.4146}{1204.4146}}]

\bibitem{Monahan:2015lha}
C.~Monahan and K.~Orginos, \emph{Phys.~Rev.~D} {\bf 91} (2015) 074513
[{\tt \href{http://arxiv.org/abs/1501.05348}{1501.05348}}]	

\end{footnotesize}
\end{thebibliography}
\end{document}